\documentclass[aps,pra,reprint,superscriptaddress]{revtex4-2}

\usepackage{amsmath,mathtools,amsfonts,stmaryrd,amssymb,dsfont} % Math packages
\usepackage[hidelinks]{hyperref}

\DeclareSymbolFont{matha}{OML}{txmi}{m}{it}% txfonts
\DeclareMathSymbol{\varv}{\mathord}{matha}{118}

\begin{document}

%\preprint{}

\title{Spin manipulation and nuclear polarization enhancement in particle beams with static magnetic fields}


\author{C.~S.~Kannis}
\email{c.kannis@fz-juelich.de}                           
\affiliation{Institut f{\"u}r Laser- und Plasmaphysik, Heinrich-Heine-Universit{\"a}t D{\"u}sseldorf, 40225 D{\"u}sseldorf, Germany}

\author{R.~Engels}    
\email{r.w.engels@fz-juelich.de}                                               
\affiliation{Insitut f{\"u}r Kernphysik, Forschungszentrum J{\"u}lich, 52428 J{\"u}lich, Germany}
\affiliation{GSI Helmholtzzentrum f{\"u}r Schwerionenforschung, 64291 Darmstadt, Germany}

\author{T.~El-Kordy}
\affiliation{FH Aachen - University of Applied Sciences, 52066 Aachen, Germany}
\affiliation{Institute of Technology and Engineering, Forschungszentrum J{\"u}lich, 52428 J{\"u}lich, Germany}

\author{N.~Faatz}
\affiliation{Insitut f{\"u}r Kernphysik, Forschungszentrum J{\"u}lich, 52428 J{\"u}lich, Germany}
\affiliation{GSI Helmholtzzentrum f{\"u}r Schwerionenforschung, 64291 Darmstadt, Germany}
\affiliation{III.~Physikalisches Institut B, RWTH Aachen University, 52062 Aachen, Germany}

\author{S.~J.~P{\"u}tz}
\affiliation{Insitut f{\"u}r Kernphysik, Forschungszentrum J{\"u}lich, 52428 J{\"u}lich, Germany}
\affiliation{GSI Helmholtzzentrum f{\"u}r Schwerionenforschung, 64291 Darmstadt, Germany}
\affiliation{Institut f{\"u}r Kernphysik, Universit{\"a}t zu K{\"o}ln, 50937 Köln, Germany}

\author{V.~Verhoeven}
\affiliation{Institut f{\"u}r Experimentalphysik I, Ruhr-Universit{\"a}t-Bochum, 44801 Bochum, Germany}

\author{T.~P.~Rakitzis}
\affiliation{Department of Physics, University of Crete, 70013 Heraklion-Crete, Greece}
\affiliation{Institute of Electronic Structure and Lasers, Foundation for Research and Technology-Hellas, 71110 Heraklion-Crete, Greece}

\author{M.~B{\"u}scher}
\email{m.buescher@fz-juelich.de}
\affiliation{Institut f{\"u}r Laser- und Plasmaphysik, Heinrich-Heine-Universit{\"a}t D{\"u}sseldorf, 40225 D{\"u}sseldorf, Germany}
\affiliation{Peter-Gr{\"u}nberg Institut, Forschungszentrum J{\"u}lich, 52428 J{\"u}lich, Germany}

\date{\today}

\begin{abstract}
	A theoretical study of spin dynamics in non-relativistic particle beams with interacting angular momenta traversing static, spatially varying magnetic fields is presented. The computational framework evaluates sinusoidal magnetic field configurations, calculating key observables such as average spin projections and state populations during the interaction. It is demonstrated that such fields can effectively enhance nuclear polarization in partially, incoherently polarized hydrogen and deuterium atomic beams, as well as coherently rotationally state-selected hydrogen deuteride molecular beams. This enhancement is attributed to transitions induced within the hyperfine regime of these systems. The study spans frequency ranges from GHz scales for atoms to hundreds of kHz for molecules, corresponding to magnetic field variations on spatial scales from submillimeters to meters.
\end{abstract}

%\keywords{}

\maketitle

\section{Introduction\label{intro}}

Control over angular momentum degrees of freedom in physical systems has enabled numerous applications across various fields. For instance, nuclear magnetic resonance (NMR) spectroscopy and magnetic resonance imaging (MRI) utilize hyperpolarized tracers to enhance signal strength, crucial for medical diagnostic accuracy~\cite{letertre2021, jorgensen2022, eills2023}. In fundamental physics research, applications such as quantum computation~\cite{neil1997}, electric dipole moment searches~\cite{harris1999}, and neutron decay studies~\cite{mendenhall2013} rely heavily on achieving high degrees of spin angular momentum polarization; hereafter referred to as spin polarization or simply polarization. While nuclear spin is often the main focus, electron spin polarization and rotational polarization are also crucial in fields like magnetometry~\cite{spiliotis2021} and molecular spectroscopy~\cite{koch2019}, respectively.

An important application of nuclear polarization is polarized fusion~\cite{kulsrud1982}. It has been shown that certain reactions, commonly referred to as five-nucleon reactions, employed for artificial fusion offer several advantages~\cite{ciullo2016, heidbrink2024} when the reactants are polarized compared to the unpolarized case. The alignment of nuclear spins in these reactions can enhance the fusion cross-section and improve reaction dynamics, leading to increased efficiency. In particular, spin polarization in the deuterium-tritium reaction has been demonstrated to enhance tritium burn efficiency, thereby reducing tritium startup inventory requirements~\cite{parisi2024}. Aneutronic fusion reactions between proton and boron (\textsuperscript{11}B) show similar advantages~\cite{ahmed2014}. Consequently, polarization could potentially play a crucial role in the development of clean energy production technologies. 

The particles involved in five-nucleon reactions have spin quantum numbers $1/2$ (tritons/helions) and $1$ (deuterons). However, tritium is scarce, radioactive, and thus expensive and difficult to handle. Helium-3 is also rare but non-radioactive and comparatively more cost-effective than tritium. As a result, research on the production of polarized fuel often focuses on hydrogen (also spin $1/2$) and deuterium atoms. These atoms and their molecules are the subject of this study.

Specifically, we investigate the dynamics of interacting angular momenta in inhomogeneous magnetic fields, considering specific initial preparation scenarios. While the preparation methods are not discussed in detail, some representative examples are mentioned. The main focus of this work is to introduce the theoretical framework for evaluating spin dynamics and to demonstrate its application to selected atomic and molecular states. Through these examples, the average spin orientation and its correlation to the populations of the related energy levels are discussed. The selected cases are intentionally chosen to highlight the enhancement of nuclear polarization achieved through carefully configured magnetic fields.

Magnetic field configurations with the analyzed characteristics can be implemented in experimental setups and serve as tools to manipulate polarization or transfer it, e.g., from the electron to the nucleus, more efficiently. The computational framework presented here can predict changes in the average polarization for experiments where particle beams pass through apparatuses that produce unequal or inhomogeneous magnetic fields. This is possible because the time-dependent equations governing spin dynamics are solved using numerical approaches.

\section{Theoretical Background\label{theory}}

The atomic and molecular systems considered here consist of at least two angular momentum entities. We denote the nuclear spin operator as $\mathbf{I}$, the total electron angular momentum operator as $\mathbf{S}$, and the molecular rotational angular momentum operator as $\mathbf{J}$. For the representation of operators and the calculation of observables, two bases are employed: the coupled and the uncoupled basis.

Assuming a system of two interacting angular momenta $\mathbf{L_{1,2}}$, the uncoupled basis vectors are defined as $\left\vert L_{1}\, l_{1} , L_{2} \, l_{2}\right\rangle = \left\vert L_{1}\, l_{1}\right\rangle \otimes \left\vert L_{2}\, l_{2}\right\rangle$, where $L_{1,2}$ are the angular momentum quantum numbers and $l_{1,2}$ are the corresponding projections along the quantization axis. In the following framework, the individual angular momentum quantum numbers $L_{1}$, $L_{2}$ are preserved; hence, it is convenient to simplify the uncoupled notation to include only the projections, i.e., $\left\vert l_{1} ,  l_{2}\right\rangle$. 

For the coupled basis, an additional angular momentum operator is defined as $\mathbf{K} = \mathbf{L_1} + \mathbf{L_2}$. This can be expressed as $\mathbf{K} = \mathbf{L_1} \otimes \mathds{1} + \mathds{1}\otimes\mathbf{L_2}$, where the operator $\mathds{1}$ in the first (second) term stands for the identity operator in the second (first) angular momentum space. The state vectors in this basis are represented by the total angular momentum quantum number $K$ and its projection $k$ along the quantization axis, denoted as $\left\vert K, k \right\rangle$. The uncoupled and coupled representations are related by a linear transformation, which will be specified whenever necessary in the studied cases.%maybe add e.g. and provide a reference to appendix

The collection of non-interacting identical systems (e.g., hydrogen atoms in a particle beam), whose observables are examined here, are assumed to be non-identically prepared. Specifically, the initial state of such an ensemble is described by a statistical mixture of the states of its constituents. For this purpose, we adopt the density operator formalism, where the (Hermitian) density operator and its matrix representation are denoted by $\rho$. The diagonal elements of the matrix are non-negative real numbers that sum to one and are referred to as populations. The off-diagonal elements are complex numbers, named coherences. The time evolution of $\rho$ is governed by the Liouville-von Neumann equation: 
\begin{equation}\label{eq:vonN}
	i\hbar\frac{\partial \rho}{\partial t} = [H, \rho] ,
\end{equation}
where $H$ is the Hamiltonian of the system. It consists of the effective hyperfine Hamiltonian $H_{0}$ and the interaction term $H_{B}$, accounting for an external magnetic field. The first term, $H_{0}$, is time-independent, while the second term, $H_B$, describes the interaction with an external sinusoidal magnetic field, defined as $B_{z} = B_{0} \sin\big(\frac{2 \pi z}{\lambda}\big)$ with $B_{0}$ the magnetic field amplitude and $\lambda$ the wavelength of the trigonometric function.

Our focus is on particles propagating parallel to the $z$-axis with a constant, non-relativistic velocity $v$ as they pass through the external magnetic field. Assuming an axially symmetric field, and applying Gauss's law for magnetism in cylindrical coordinates $(r, \phi, z)$, we obtain the radial component $B_{r} = -\frac{r}{2} \frac{\partial B_{z}}{\partial z} = - B_{0} \frac{\pi r}{\lambda}\cos\big(\frac{2 \pi z}{\lambda}\big)$, where $r$ is the radial distance from the $z$-axis. Thus, for a particle traveling along the $z$-axis, i.e., at $r=0$, only the longitudinal component $B_{z}$ is present.

\begin{figure}
	%\centering
	\includegraphics[width=1.0\columnwidth]{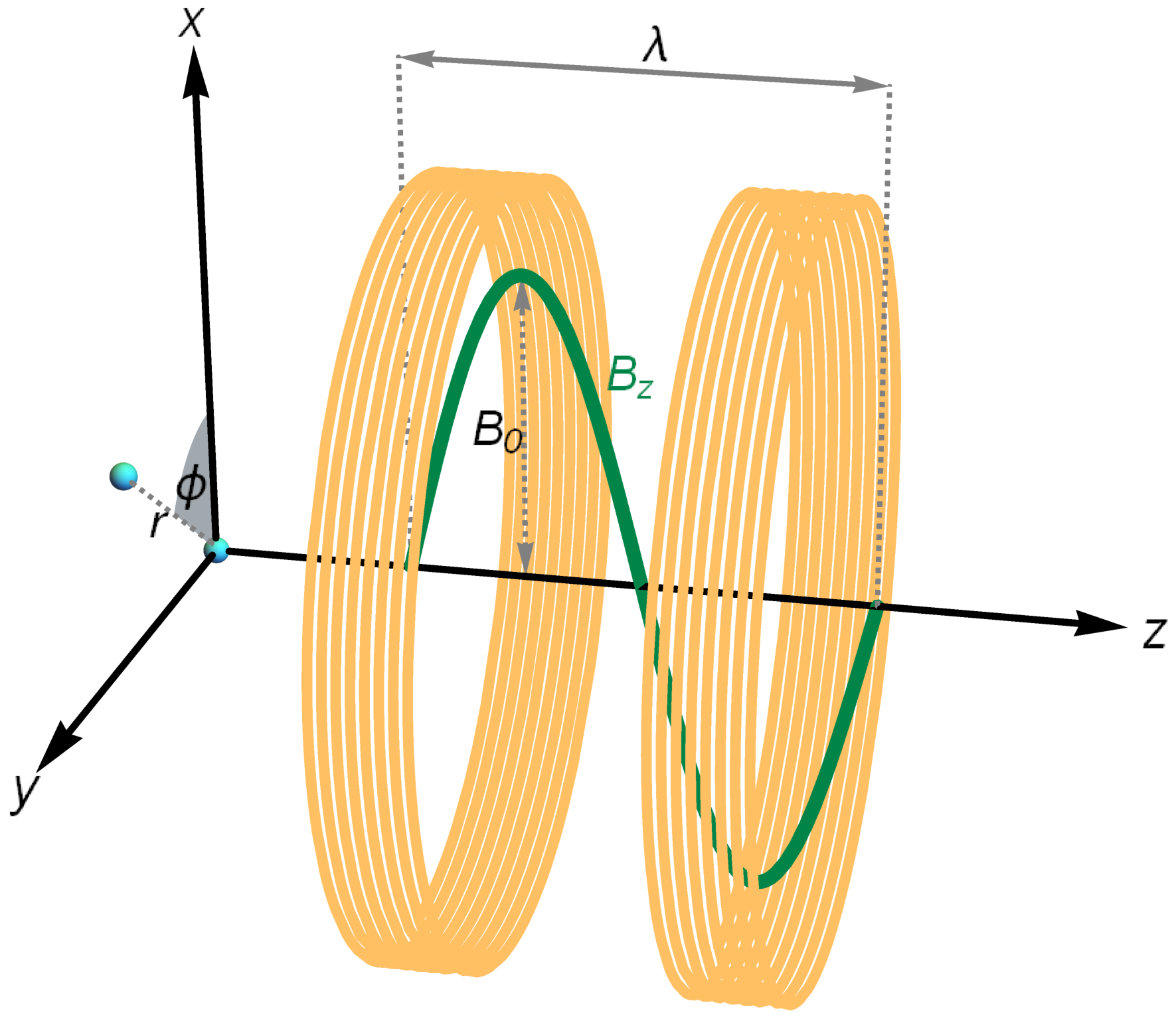}
	\caption{Schematic of two particles (cyan spheres), one at $r=0$ and one at $r\neq0$, on the $z=0$ plane in the lab frame, in the presence of a spatially varying magnetic field. The orange circles represent two cylindrical coils generating a longitudinal field $B_{z}$, along with a radial component $B_{r}$ (omitted for simplicity). For illustration purposes, the magnetic field configuration has been shifted to the right.\label{figm1}}
\end{figure}

The description above corresponds to the laboratory reference frame (see Fig.~\ref{figm1}), where the two key features are: (a) time-dependent particle position $(r, \phi, z)$ and (b) static--but spatially varying--magnetic fields. The transverse coordinates $(r, \phi)$ are assumed to remain constant throughout the particle's motion, while the longitudinal position $z$ evolves as $z = v t$. It is convenient to change the reference frame and employ the rest frame of the particle. In this frame, the particle's position is constant $(r^\prime = r, \phi^\prime = \phi, z^\prime=0)$, where $z^\prime=0$ by convention. Since the transverse coordinates remain unaffected by the change of reference frame, we will use unprimed coordinates in the rest frame; the distinction between the two frames becomes clear from the context.

As a result of this transformation, the spatial variation of the magnetic field components in the lab frame is perceived as a time-dependent variation in the particle's rest frame. The magnetic field in this frame is obtained by substituting $z$ with $v t$ in the lab-frame expressions, yielding
\begin{equation}\label{eq:bfieldt}
	\mathbf{B} = B_{z} \hat{\mathbf{z}}  + B_{r} \hat{\mathbf{r}} = B_{0} \sin\Big(\frac{2 \pi v t}{\lambda}\Big)  \hat{\mathbf{z}} - B_{0} \frac{\pi r}{\lambda}\cos\Big(\frac{2 \pi v t}{\lambda}\Big) \hat{\mathbf{r}},
\end{equation}
or equivalently, in Cartesian coordinates: $\mathbf{B} = B_{r} \cos\phi \, \hat{\mathbf{x}} \, + \, B_{r} \sin\phi \, \hat{\mathbf{y}} \, + \,  B_{z} \, \hat{\mathbf{z}}$, where the azimuthal angle $\phi$ is measured from the $x$-axis.

In the rest frame, the key features are now: (a) constant particle position and (b) time-dependent magnetic field components. Consequently, there is no need to include a kinetic term in the Hamiltonian, and the angular momenta involved in the particle dynamics can be treated as if the particle were stationary in a time-dependent magnetic field given by Eq.~(\ref{eq:bfieldt}). This equivalent description will be employed to evaluate the systems' dynamics. It is a purely magnetic model that neglects the presence of additional external fields (see Sec.~\ref{limitations} for a discussion of electric fields), ensuring that the angular momentum quantum numbers $I$, $S$, and $J$ remain conserved.

\begin{figure}
	%\centering
	\includegraphics[width=1.0\columnwidth]{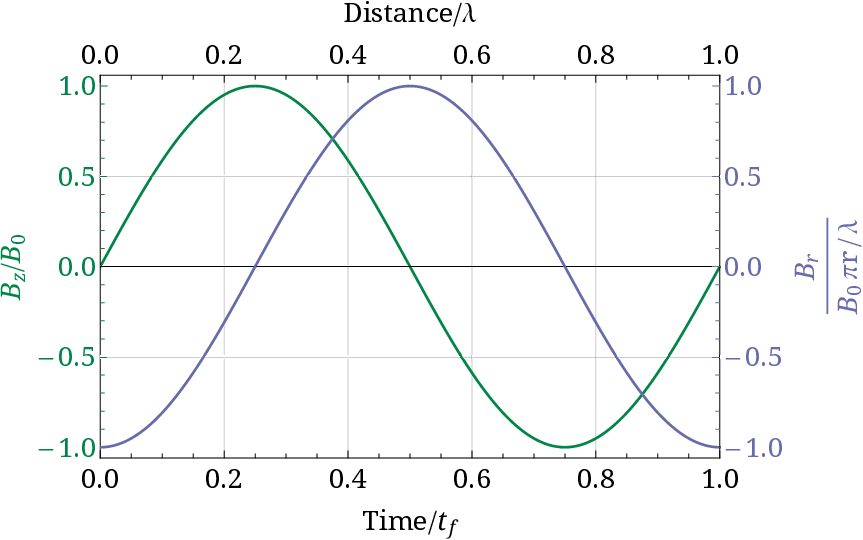}
	\caption{Longitudinal and radial magnetic field components from Eq.~(\ref{eq:bfieldt}). The top horizontal axis represents the spatial variation of the fields in the laboratory frame, while the bottom horizontal axis corresponds to the time-dependent variation experienced by a moving particle.\label{fig0}}
\end{figure}

%From the above, it is apparent that our framework can be applied not only to particles moving through a static magnetic field but also to particles at rest experiencing only a time dependent magnetic field of the form given in Eq.~(\ref{eq:bfieldt}). However, our group's experimental focus is primarily on particle beams; therefore, this is the main context of our analysis.

It is useful to introduce the time of flight $t_{f} = \lambda / v$, which is defined to be equal to the period of the trigonometric function. In other words, we choose to determine the optimal magnetic field configurations based on the corresponding time of flight. This is the key parameter in our analysis, with residual parameters, such as $\lambda$, determined by it and the experimental conditions. In the following section, the spin dynamics are obtained by solving Eq.~(\ref{eq:vonN}), and relevant observables are calculated for time $t$, $0 \leq t \leq t_{f}$. The longitudinal and radial magnetic field components from Eq.~(\ref{eq:bfieldt}) are illustrated in Fig.~\ref{fig0}, demonstrating the equivalence between the spatial variation of the fields in the laboratory frame and the time-dependent variation in the particle's frame. As a convention, the quantization axis is taken to be fixed and aligned with the $z$-axis. Positive spin projections are therefore parallel to the first (positive-valued) half of the longitudinal magnetic field component, while negative projections align with the second half.

\section{Results and Discussion\label{resndisc}}

This section is divided into three subsections. First, we analyze the spin dynamics for hydrogen and deuterium atoms, then for HD molecules, and finally, we discuss limitations of the analysis.

\subsection{Hydrogen (H) and deuterium (D) atoms}

We focus on atomic states where the total electron angular momentum equals the electron spin, thereby avoiding unnecessary complexity introduced by additional angular momenta. Specifically, we examine the ground and first metastable states of hydrogen (H) and deuterium (D) atoms. The effective hyperfine Hamiltonian governing such states is expressed as~\cite{bethe1957}:
\begin{equation}\label{eq:h0}
	H_0 = \frac{A}{{\hbar}^2} \mathbf{I} \cdot \mathbf{S} ,
\end{equation}
where $A$ is the hyperfine-structure constant. The electron spin is $1/2$, while the nuclear spin is $1/2$ for H and $1$ for D. The interaction with the applied magnetic field $\mathbf{B}$ is described by
\begin{equation}\label{eq:hb}
	H_{B} = -\boldsymbol{\mu} \cdot \mathbf{B} = - \Big( \frac{g_{s} \mu_{B}}{\hbar} \mathbf{S} + \frac{g_{I} \mu_{N}}{\hbar} \mathbf{I} \Big) \cdot \mathbf{B} ,
\end{equation}
where $\boldsymbol{\mu}$ is the atomic magnetic dipole moment, comprising contributions from both the electron and nuclear magnetic moments. These contributions are proportional to the g-factors $g_{s, I}$ and the Bohr and nuclear magnetons, $\mu_{B, N}$, respectively. In SI units, the latter are given by  $\mu_{B} = 9.27\times 10^{-24}$~$\text{J/T}$ and $\mu_{N} = 5.05\times 10^{-27}$~$\text{J/T}$. The g-factors are well approximated by their free-particle's values: $g_{s} = -2.002$ for an electron (no orbital contribution), $g_{I}=5.586$ for a proton (H nucleus) and $g_{I}=0.857$ for a deuteron (D nucleus).

Before presenting the results for the atomic systems, it is useful to introduce one more observable related to spin polarization, which is calculated using the density matrix: the average spin projection of the nucleus ($\left\langle m_{q, n} \right\rangle$) and the electron ($\left\langle m_{q, e} \right\rangle$), defined as
\begin{equation}\label{eq:spin}
	\left\langle m_{q, (n/e)} \right\rangle = \frac{1}{\hbar}\left\langle	S_{q, (n/e)}  \right\rangle = \frac{1}{\hbar} Tr(\rho	S_{q, (n/e)} )
\end{equation}
along the $q$-th axis ($x$, $y$, or $z$). $S_{q}$'s denote the spin matrices and the subscripts $n$ and $e$ correspond to the nuclear and electron spins, respectively. The spin matrices in the uncoupled representation can be derived from the Pauli matrices and are listed in App.~\ref{app1}, along with all basis states for the two representations and the corresponding transformation. As expected, the spin projection along a given axis relates to the (vector) polarization along the same axis, by dividing the former with the corresponding quantum number. For example, the electron and proton spin polarizations are twice the corresponding spin projections, whereas the deuteron polarization equals its spin projection.

In our first example, for H, we consider an initial equal population of the states $|F=1, m_F = 1 \rangle$ and $|F=1 , m_F =0 \rangle$ with the quantization axis parallel to the $z$-axis. This is described by the density matrix $\rho_{c}=\text{diag}(1/2,1/2,0,0)$, whose off-diagonal elements vanish, and can hence be characterized as incoherent. Such an initial preparation can be achieved for the metastable $2 S_{1/2}$ state of H~\cite{cesati1966}, for instance, by directing an unpolarized beam into a region with a magnetic field of $\sim 57.5$ mT and a weak transverse electric field. In this setup, two of the states are quenched to the ground state $1 S_{1/2}$, and the beam is then gradually/adiabatically transferred to a low-field region $B \rightarrow 0$, where the eigenstates are described by the coupled basis. For the ground state of H~\cite{kponou2008}, optically pumped polarized rubidium vapor can induce electron spin polarization in a high magnetic field, which is subsequently adiabatically reduced to a value well below the critical field~\cite{steffens2003}.

\begin{figure}
	%\centering
	\includegraphics[width=1.0\columnwidth]{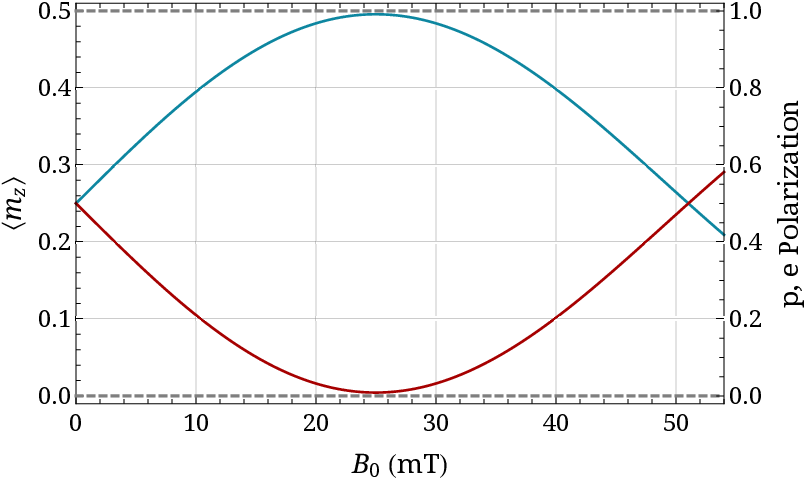}
	\caption{Average spin projections $\left\langle m_{z, p}\right\rangle$ (blue) and $\left\langle m_{z, e}\right\rangle$ (red) of a ground-state H atom at time $t_{f}\sim 0.7$~ns as a function of $B_0$, for an initially incoherently, partially polarized beam at $r=0$. The right vertical axis represents the corresponding polarization.\label{fig1}}
\end{figure}

Independent of the preparation method, the resulting beam exhibits proton and electron spin polarizations of $50\%$, which persist in the absence of external fields~\footnote{Practically, a small, well-controlled magnetic field aligned with the polarization axis (quantization axis) is often applied to minimize unwanted spin precession}. Now, consider this beam passing through the static sinusoidal field, introduced earlier, at $r = 0$, so that it experiences only the $B_{z}$ component. By choosing the field wavelength $\lambda$ so that the time of flight matches the inverse of the hyperfine frequency $\varv_{HF} = A/h$, i.e., $t_{f} = 1/\varv_{HF}$, it becomes possible to maximize proton polarization. Figure~\ref{fig1} shows the average spin projection along the $z$-axis for the proton ($\left\langle m_{z, p}\right\rangle$) and electron ($\left\langle m_{z, e}\right\rangle$) in a ground-state atom at time $t_{f}$ as a function of the magnetic field amplitude $B_{0}$. The hyperfine frequency is $\varv_{HF} = 1.42$~GHz~\cite{diermaier2017}, or equivalently $A = h \varv_{HF} = 5.87$~{\textmu eV}.

\begin{figure}
	\includegraphics[width=1.0\columnwidth]{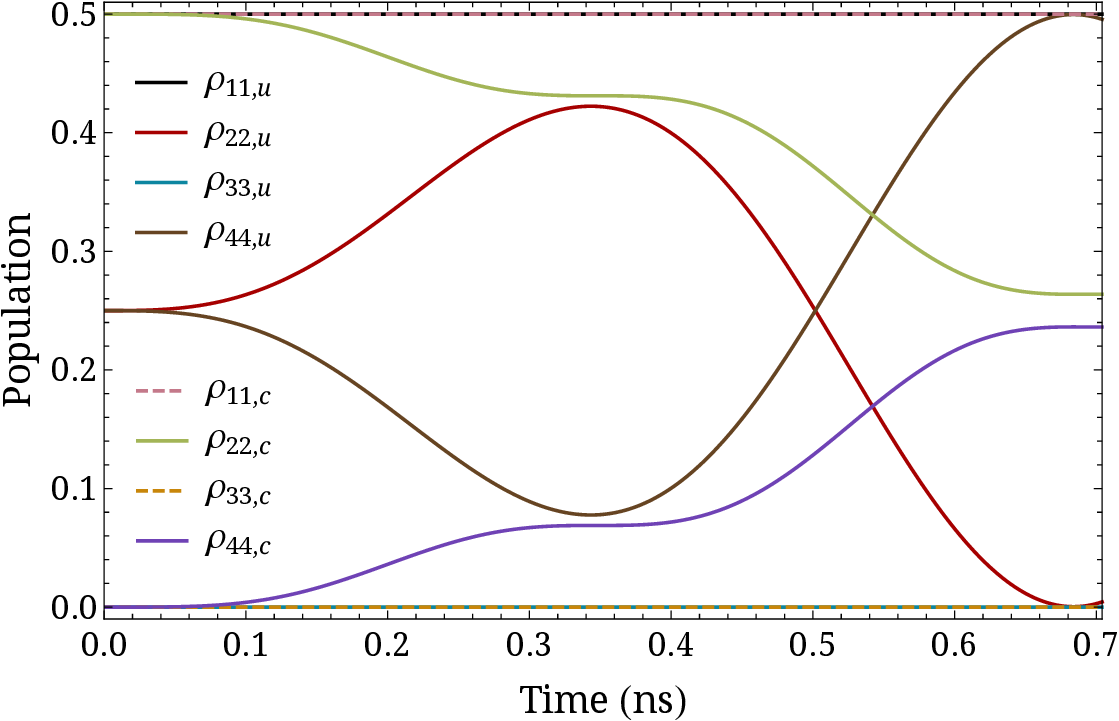}
	\caption{Populations of uncoupled (subscript $u$) and coupled (subscript $c$) states during the motion through the applied longitudinal magnetic field with $v/\lambda = 1.42$~GHz and $B_{0} = 25$~mT. The states are ordered according to App.~\ref{app1a}.\label{fig2}}
\end{figure}

It is insightful to look closer at the dynamics during passage through the sinusoidal field for $B_{0} = 25$~mT, at which the nuclear polarization peaks at $99.12\%$. Figure~\ref{fig2} illustrates the populations in terms of the coupled and uncoupled states. The two states that transform trivially (overlapping lines $\rho_{1 1,u} {=} \rho_{1 1,c}$ and $\rho_{3 3,u} {=} \rho_{3 3,c}$) have constant populations over time. The second and fourth state of the uncoupled representation directly reveal the evolution of the average spin projection of the electron and the proton, respectively. This occurs because the other states have constant populations, and the longitudinal spin matrix is diagonal in this representation. Measurements of occupation in the uncoupled basis therefore provide direct information about the spin polarization. On the other hand, in the coupled basis, this information is encoded in a more indirect manner (see off-diagonals in Eq.~(\ref{eq:ex1})). No spin polarization develops along the $x$- or $y$-axes.

Obtaining the eigenenergies, and subsequently the evolution dynamics for the applied conditions, involves the evaluation of elliptical integrals, which is performed numerically. The results reveal oscillations between states with the same total angular momentum projection, $ m_{F}  =  m_{S}  +  m_{I}$, when their initial populations are uneven. This behavior reflects the selection rule $\Delta m_{F} = 0$ for magnetic dipole transitions, indicating that the sum of spin projections along the $z$-axis, or equivalently the total longitudinal polarization, is preserved.

The experimental realization of this field for a beam with a kinetic energy of $1$~keV (velocity $v\sim 4.4\times 10^{5}$~m/s) corresponds to a wavelength of $\lambda\sim 0.3$~mm. This wavelength increases proportionally with the beam velocity, as the time of flight $t_{f}$ is fixed to the inverse of the hyperfine frequency. Achieving submillimeter periodic magnetic fields with amplitudes of several millitesla cannot be accomplished using conventional wire coils. Instead, techniques such as microfabricated magnetic arrays, superconducting microcoils, magnetized ferromagnetic gratings, and other advanced methods are required. For metastable $2S_{1/2}$ H, where the hyperfine interaction is approximately eight times weaker, $A/h = 177.56$~MHz~\cite{kolachevsky2009}, the time of flight must increase by the same factor, resulting in a wavelength of $\lambda\sim 2.5$~mm under the same beam conditions. As expected, the proton polarization in the metastable state peaks at $B_{0}\sim 3.1$~mT, namely at a field that is eight times weaker. This effect can be observed via Lamb-shift polarimetry~\cite{engels2003}.

\begin{figure}
	\includegraphics[width=1.0\columnwidth]{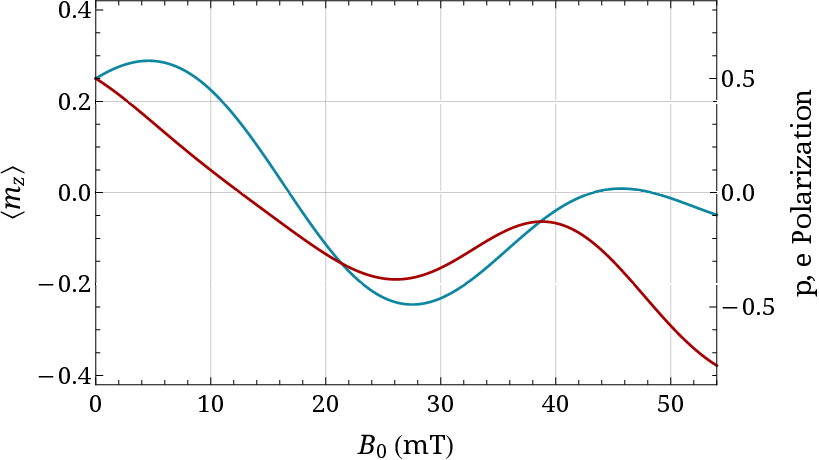}
	\caption{Average spin projections $\left\langle m_{z, p}\right\rangle$ (blue) and $\left\langle m_{z, e}\right\rangle$ (red) of a ground-state H atom at time $t_{f}\sim 0.7$~ns as a function of $B_0$, for an initially incoherently, partially polarized beam at $r=0.3$~mm. The right vertical axis represents the corresponding polarization.\label{fig3}}
\end{figure}

\begin{figure}
	\includegraphics[width=1.0\columnwidth]{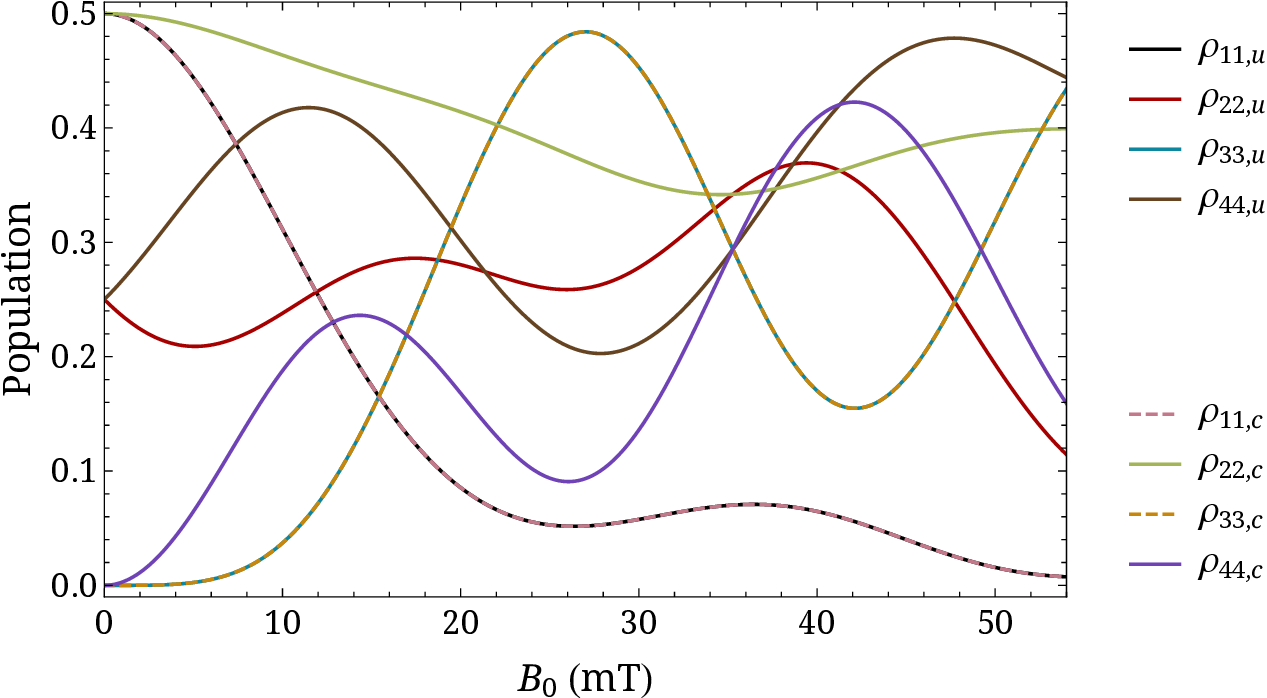}
	\caption{Populations of uncoupled ($u$) and coupled ($c$) states at time $t_{f}\sim 0.7$~ns as a function of $B_0$, at $r=0.3$~mm. The states are ordered according to App.~\ref{app1a}.\label{fig4}}
\end{figure}

For particles positioned off-axis, $r\neq 0$, the additional radial field component $B_{r}$ disrupts the polarization dynamics. The magnitude of $B_{r}$ is proportional to the ratio $r/\lambda$, as indicated by Eq.~(\ref{eq:bfieldt}). At $r/\lambda\ll 1$, the influence of $B_{r}$ is negligible; for instance, at $r/\lambda=10^{-3}$, the loss of polarization is well below $1\%$ for the magnetic field amplitudes considered here. As the radial distance increases, the impact of $B_{r}$ becomes significant, leading to differing polarization results across radii. Figure~\ref{fig3} shows an example of this effect for $r=\lambda=0.3$~mm. In this case, polarization along the quantization axis is disrupted, and a transverse polarization component is developed due to the radial field. Consequently, the sum of average spin projections is no longer conserved. In contrast, the sum of the populations remains constant, as illustrated in Fig.~\ref{fig4}. The Hamiltonian matrix governing such dynamics is derived in~\cite{kannis2023} and presented for both representations of H and D atoms. The size of the matrix is $(2 S + 1) (2 I + 1) \times (2 S + 1) (2 I + 1)$, leading to a system of $(2 S + 1)^2 (2 I + 1)^2$ differential equations for the corresponding density matrix elements, as described by Eq.~(\ref{eq:vonN}). Due to their large number and relatively complex form, the explicit equations used to obtain the presented results are omitted.

This effect has also been experimentally observed~\cite{engels2021} for metastable beams initially prepared in a single hyperfine state during the investigations of Sona transitions~\cite{sona1967}. Sona transition units are components that generate a longitudinal magnetic field with a reversing direction, thereby prompting more extensive studies on interacting angular momenta in such environments. A detailed discussion of this phenomenon is beyond the scope of this work; further experimental investigations will be carried out in the future. Briefly, similar to $B_{z}$, the $B_{r}$ component induces magnetic dipole transitions, but with the selection rule $\Delta m_{F} = \pm 1$.

An alternative method for preparing initial spin states, based on molecular photodissociation~\cite{rakitzis2003,spiliotis2021b}, can lead to up to $100\%$ nuclear polarization. This approach uses very short laser pulses ($<300$~ps) to dissociate hydrogen halides, transferring the laser circular polarization to the electrons of the fragments ($\hbar/2$ to the H-atom electron and $\hbar/2$ to the valence electrons of the halide atom), while leaving the nuclear spin effectively unpolarized. The resulting state for H atoms is described by the density matrix $\rho_{u}=\text{diag}(1/2,1/2,0,0)$ for positive electron polarization. Transforming this to the coupled basis yields $\rho_{c}= \mathcal{Q} \rho_{u}\mathcal{Q}^{-1}$, which includes two non-zero off-diagonal elements, allowing polarization to oscillate between the electron and nuclear spins with a period of $1/\varv_{HF}$. This is commonly referred to as coherent preparation, which in terms of timescales differs significantly from the incoherent preparation discussed earlier. 

For incoherent preparation, the initial polarization is also induced in the electron spin, but within a hyperfine-resolved regime at high magnetic fields. This means that the atomic states approach the uncoupled states, but after reducing the magnetic field adiabatically to the low-field limit, the atomic states transition to the coupled states. This slow process results in a partial, steady-state polarization of $50\%$ for both electron and proton spins. In contrast, photodissociation with short pulses, which operates without resolving the hyperfine structure, can induce up to $100\%$ electron spin polarization in a field-free environment. This polarization oscillates due to the hyperfine interaction, transferring entirely to the nuclear spin after a time of $1/2\varv_{HF}$. Several studies~\cite{rakitzis2004, sofikitis2018, spiliotis2021b} validate this technique, demonstrating its scalability to macroscopic quantities by means of high-power lasers.

\begin{figure}
	\includegraphics[width=1.0\columnwidth]{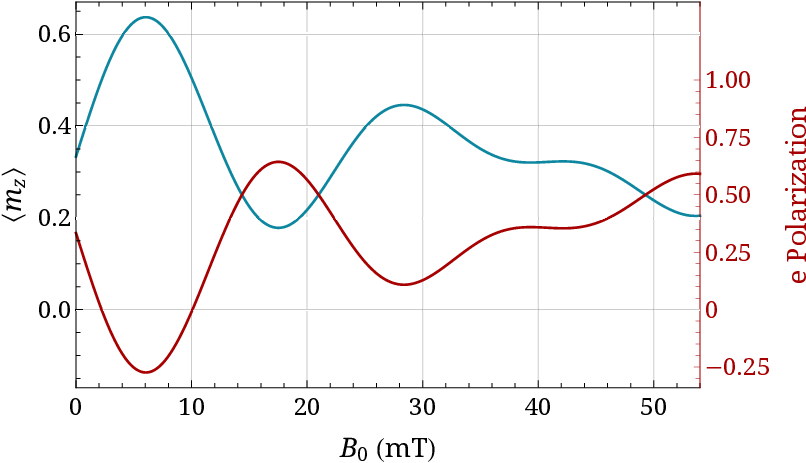}
	\caption{Average spin projections $\left\langle m_{z, d}\right\rangle$ (blue) and $\left\langle m_{z, e}\right\rangle$ (red) of a ground-state D atom at $t_{f}\sim 3$~ns as a function of $B_0$, for an initially incoherently, partially polarized beam at $r=0$. The right vertical axis represents the electron spin polarization, while the deuteron polarization is directly given by $\left\langle m_{z, d}\right\rangle$.\label{fig5}}
\end{figure}

Next, we examine a similar, incoherent preparation of deuterium atoms. Both methods discussed earlier for ground-state and metastable H can also be applied to D, e.g., for the metastable state see~\cite{cesati19662}. At $t=0$, the resulting density matrix takes the form $\rho_{c}=\text{diag}(1/3,1/3,1/3,0,0,0)$, indicating equal population of the states $\left\vert F=3/2, m_F = 3/2 \right\rangle$, $\left\vert F=3/2, m_F = 1/2 \right\rangle$, and $\left\vert F=3/2, m_F = -1/2 \right\rangle$. In contrast to H, the hyperfine frequency is not equal to $A/h$, but $\varv_{HF} = 3A/2h$, corresponding to the hyperfine splitting (in frequency units) between the quadruplet $F=3/2$ and the doublet $F=1/2$. For the ground state, this has been measured as $\varv_{HF} = 327.38$~MHz~\cite{wineland1972}. Consequently, the favorable time of flight is $t_{f} \sim 3$~ns. Figure~\ref{fig5} illustrates how the average spin projection of the electron and deuteron vary as a function of the amplitude $B_{0}$ of the longitudinal magnetic field $B_{z}$. Recall that the electron spin polarization is twice the expectation value of its spin projection, and hence both the electron and deuteron exhibit equal polarization of $1/3$ at $B_{0} = 0$.

At $B_{0} = 6$~mT, the nuclear polarization reaches a peak of $63.7\%$. This value is slightly higher than the maximum polarization of $59.3\%$ achieved by molecular photodissociation~\cite{sofikitis2017}, which involves coherent preparation of electron spin polarization followed by polarization transfer via the hyperfine interaction in the absence of external fields.

\begin{figure}
	\includegraphics[width=1.0\columnwidth]{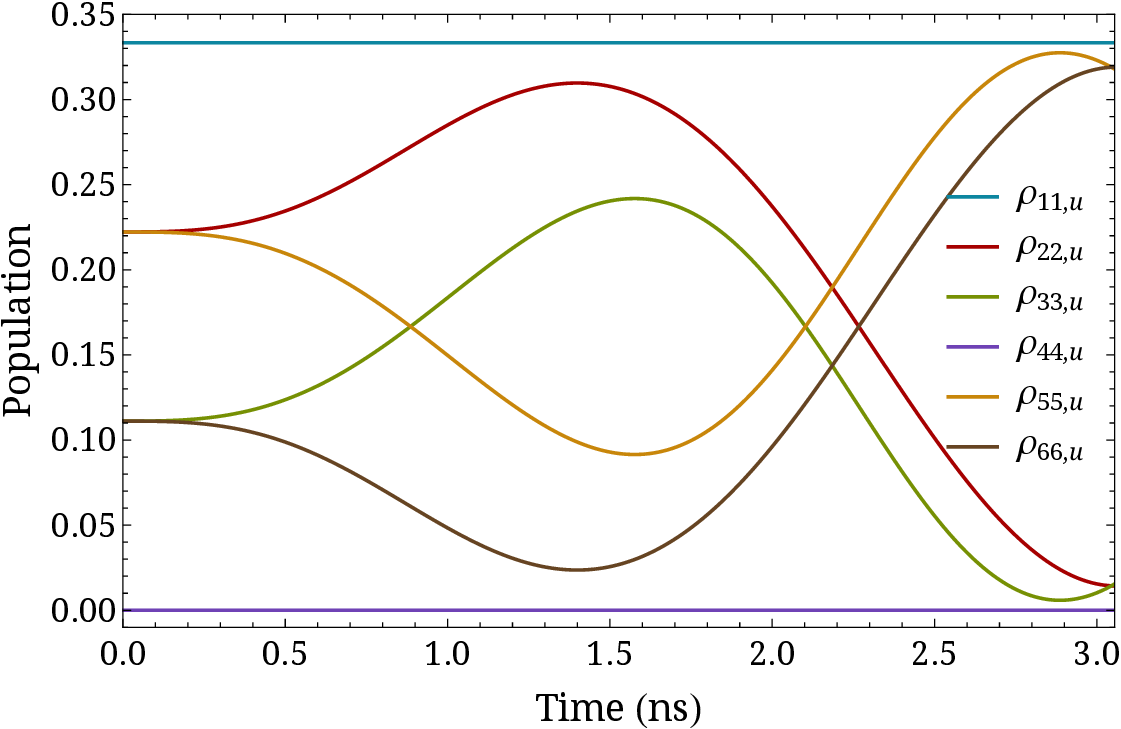}
	\caption{Populations of uncoupled states during the motion through the applied longitudinal magnetic field with amplitude $B_{0} = 6$~mT. The states are ordered according to App.~\ref{app1b}.\label{fig6}}
\end{figure}

\begin{figure}
	\includegraphics[width=1.0\columnwidth]{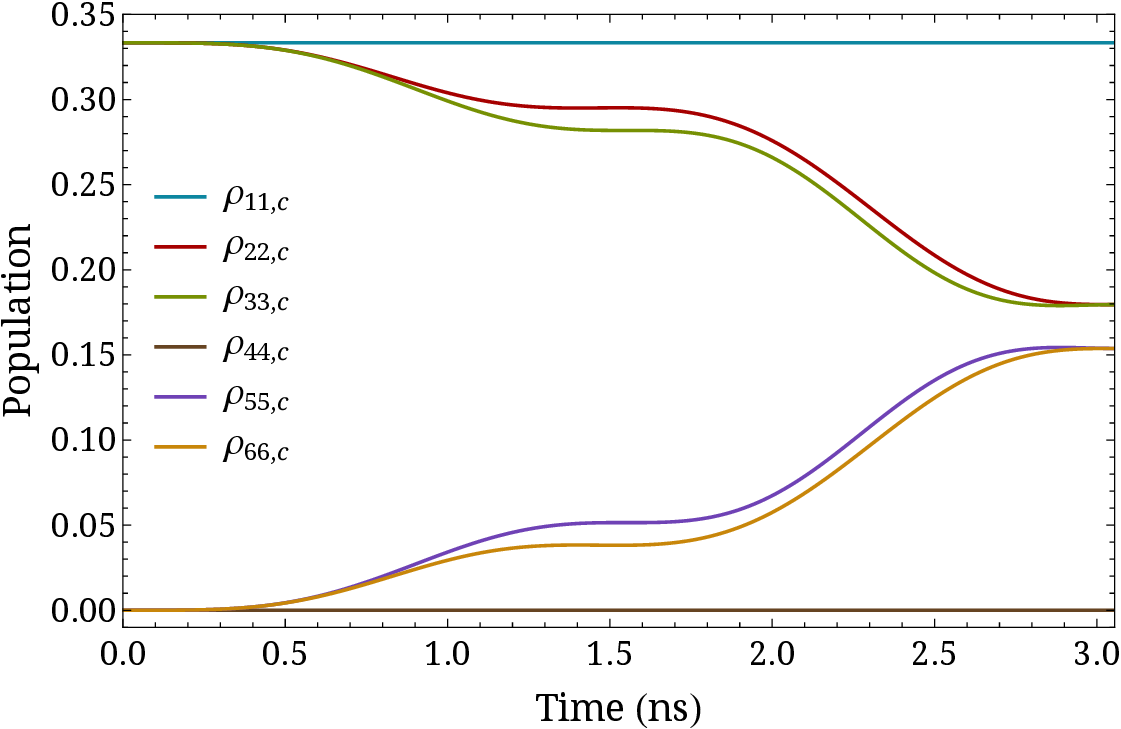}
	\caption{Populations of coupled states during the motion through the applied longitudinal magnetic field with amplitude $B_{0} = 6$~mT. The states are ordered according to App.~\ref{app1b}.\label{fig7}}
\end{figure}

For comparison with hydrogen, we plot the time evolution of the state populations in the uncoupled and coupled representation for $B_{0} = 6$~mT in Fig.~\ref{fig6} and~\ref{fig7}, respectively. The populations $\rho_{11,u}$ and $\rho_{11,c}$, as well as $\rho_{44,u}$ and $\rho_{44,c}$, correspond to the same states assigned by electron and proton spin projections parallel to each other, and parallel and antiparallel to the quantization axis, respectively; for details see App.~\ref{app1b}. Overall, the sum of average spin projections and that of the state populations are preserved in the presence of a longitudinal sinusoidal field. That is because, as in H, transitions occur obeying the selection rule $\Delta m_F = 0$. Specifically, states with the same $m_{F}$ in Fig.~\ref{fig7}, appear as mirror images of each other. This behavior is also exhibited by the uncoupled states in Fig.~\ref{fig6}, e.g., $\left\vert \frac{1}{2}, 0\right\rangle$ (state $2$) and $\left\vert -\frac{1}{2}, 1 \right\rangle$ (state $6$). No polarization buildup occurs in the transverse plane.

The realization of such a setup for a $1$-keV beam (velocity $v\sim 3.1\times 10^{5}$~m/s) corresponds to a wavelength $\lambda\sim 0.9$~mm. In the metastable $2S_{1/2}$ state, where the hyperfine splitting is $40.92$~MHz~\cite{kolachevsky2004}, achieving an eightfold longer time of flight would require a correspondingly longer wavelength, $\lambda\sim7.6$~mm. Similarly, the optimal field amplitude for high nuclear polarization would be eight times weaker, i.e., $0.75$~mT. It should be noted that particle beams with such a small diameter are generally not very intense, making these examples rather idealized cases.

\begin{figure}
	\includegraphics[width=1.0\columnwidth]{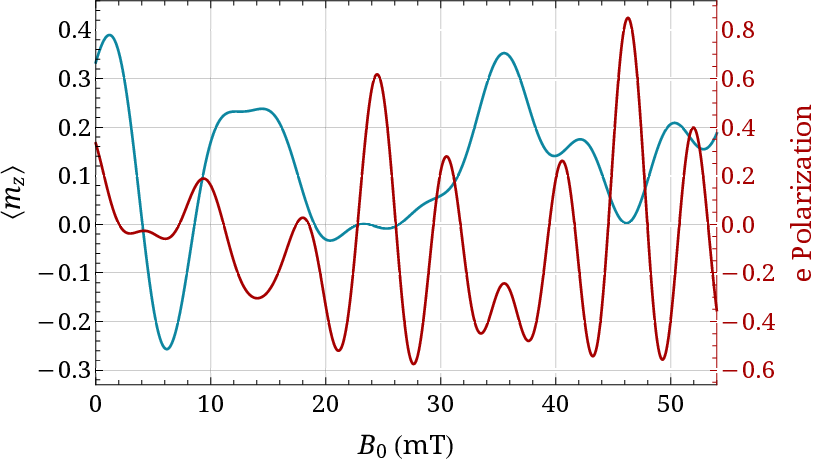}
	\caption{Average spin projections $\left\langle m_{z, d}\right\rangle$ (blue) and $\left\langle m_{z, e}\right\rangle$ (red) of a ground-state D atom at time $t_{f}\sim 3$~ns as a function of $B_0$, for an initially incoherently, partially polarized beam at $r\sim0.9$~mm. The right vertical axis represents the electron spin polarization, while the deuteron polarization is directly given by $\left\langle m_{z, d}\right\rangle$.\label{fig8}}
\end{figure}

Next, we provide an example of how the radial magnetic field component affects the spin polarization. In particular, we consider a radial distance equal to the wavelength $\lambda$ of the applied field. Figure~\ref{fig8} shows how the average spin projections along the $z$-axis vary as $B_{0}$ increases for this radial distance. The longitudinal polarization is disrupted due to additional transitions induced by the radial field component $B_{r}$, leading to nonzero transverse polarization. As a result, the total initial polarization is not conserved. However, these transitions redistribute the state populations without losses. To illustrate this behavior, graphs of the state populations at this radial distance, evaluated at the final point of the magnetic field, as a function of $B_{0}$, are provided in App.~\ref{app2a} (see Figs.~\ref{fig9} and~\ref{fig10}).

\subsection{Hydrogen deuteride (HD) molecule}

The case of two interacting angular momenta, described by a term of the form $\mathbf{I} \cdot \mathbf{S}$, in sinusoidally varying magnetic fields was reviewed in the preceding section. Naturally, the next logical step is to investigate the spin dynamics of a more complex system. Here, we focus on HD, the molecule formed by combining the previously studied atoms. The analysis considers the electronic state $^{1} \Sigma$ and the rotational levels $J=1$ and $J=2$. The effective hyperfine Hamiltonian is expressed as~\cite{ramsey1957}:
\begin{equation}\label{eq:h0hd}
	\begin{split}
		H_0 = & -\frac{c_{p}}{{\hbar}^2} \mathbf{I_{p}} \cdot \mathbf{J} - \frac{c_{d}}{{\hbar}^2} \mathbf{I_{d}} \cdot \mathbf{J} +\frac{\delta}{\hbar^{2}} \mathbf{I_{p}} \cdot \mathbf{I_{d}} \\
		& +\frac{5 d_{1}}{(2 J -1)(2 J + 3) \hbar^{4}}\Big[\frac{3}{2} (\mathbf{I_{p}} \cdot \mathbf{J})(\mathbf{I_{d}} \cdot \mathbf{J}) \\
		& + \frac{3}{2} (\mathbf{I_{d}} \cdot \mathbf{J})(\mathbf{I_{p}} \cdot \mathbf{J})- \mathbf{I_{p}} \cdot \mathbf{I_{d}} \mathbf{J}^2 \Big] \\
		& +\frac{5 d_{2}}{(2 J -1)(2 J + 3) \hbar^{4}}\Big[3 (\mathbf{I_{d}} \cdot \mathbf{J})^2 + \frac{3}{2} (\mathbf{I_{d}} \cdot \mathbf{J}) \\
		& - \mathbf{I_{d}}^2 \mathbf{J}^2 \Big] ,
	\end{split}
\end{equation}
where $\mathbf{I_{p}}$ and $\mathbf{I_{d}}$ denote the proton and deuteron spins, respectively, and $\mathbf{J}$ is the rotational angular momentum of the molecule. Detailed derivations of these interactions can be found in~\cite{ramsey1957, ramsey1953,ramsey1955}. The parameters of the interactions have been experimentally determined (in frequency units): $c_{p}/ h = 85589$~Hz, $c_{d}/h = 13118$~Hz, $\delta/h = 43$~Hz, $d_{1}/h = 17764$~Hz, and $d_{2}/h = -22452$~Hz. 

The interaction Hamiltonian with an external magnetic field $\mathbf{B}$ is given by~\cite{ramsey1957}:
\begin{equation}\label{eq:hbhd}
	\begin{split}
		H_{B} = & -\frac{a_{p}^\prime}{\hbar} \mathbf{I_{p}} \cdot \mathbf{B} -\frac{a_{d}^\prime}{\hbar} \mathbf{I_{d}} \cdot \mathbf{B}  -\frac{b^\prime}{\hbar} \mathbf{J} \cdot \mathbf{B} \\
		& -\frac{5 f^{\prime}}{3 (2 J - 1) (2 J + 3) \hbar^2} \big(3 (\mathbf{J} \cdot \mathbf{B})^2 -\mathbf{J}^2 \mathbf{B}^2 \big) ,
	\end{split}
\end{equation}
where the interaction parameters are: $a_{p}^\prime/ h = 4257.796\times 10^{4}$~Hz/T, $a_{d}^\prime/ h = 653.5832\times 10^{4}$~Hz/T, $b^\prime/ h = 505.5870\times 10^{4}$~Hz/T, and $f^\prime/ h = -2630$~Hz/T\textsuperscript{2}.

Coherent excitation techniques enable the preparation of molecules in rotational states with specific projections (e.g., $m_J = + J$)~\cite{mukherjee2010, vitanov2017}, effectively transferring $100\%$ of the population to such a state. This polarization of molecular rotational angular momentum can subsequently transfer to the nuclear spins through hyperfine interactions~\cite{altkorn1985,orrewing1994}. For ortho-species of homoatomic molecules such as H\textsubscript{2} ($J=1$) and D\textsubscript{2} ($J=2$), polarization values reach up to $99\%$~\cite{rakitzis2005} and $72\%$~\cite{bartlett2010}, respectively, under field-free conditions.

However, achieving high nuclear polarization is more challenging in some molecules, especially heteroatomic ones with at least two nonzero nuclear spins, such as HD~\cite{bartlett2009}. For $J=1$, proton and deuteron polarizations peak at $70\%$ and $64\%$, respectively, within $50$~\textmu{s} after coherent state preparation, in the absence of external fields. For $J=2$, these values are approximately $64-65\%$. Homogeneous magnetic fields, which decouple interacting angular momenta~\cite{kannis2018}, cannot enhance nuclear polarization.  Instead, the sinusoidal-field technique presented earlier for atomic systems is adapted here to enhance nuclear polarization in coherent rotationally state-selected molecules. The main challenge lies in determining the optimal time of flight, as molecular hyperfine interactions are more complex than those in atomic systems. Nonetheless, the approach remains consistent: the time of flight is chosen based on the size of hyperfine splittings. For HD, these are on the order of tens to hundreds of kHz, significantly smaller than the GHz-scale splittings in atomic systems. 

\begin{figure}
	\includegraphics[width=1.0\columnwidth]{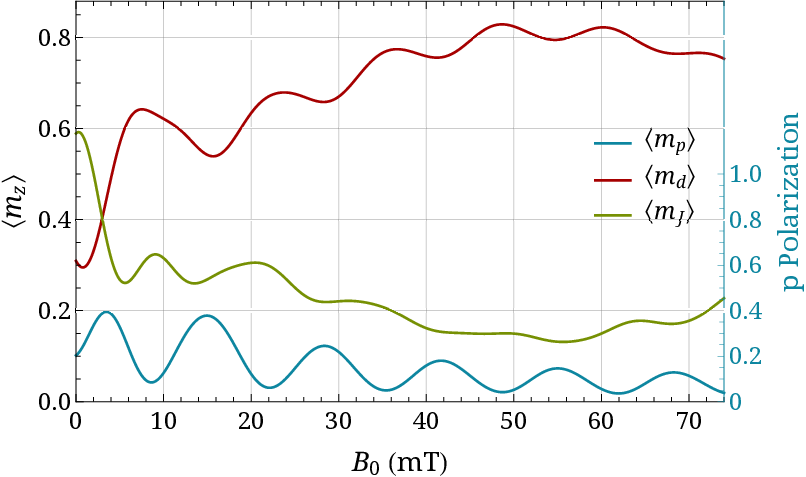}
	\caption{Average spin projections $\left\langle m_{p}\right\rangle$, $\left\langle m_{d}\right\rangle$, and $\left\langle m_{J}\right\rangle$ along the $z$-axis for $J = 1$ at $t_{f}\sim 6.5$~\textmu{s} as a function of $B_0$. The right vertical axis represents the proton spin polarization, while the deuteron polarization is directly given by $\left\langle m_{d}\right\rangle$. An initially coherently rotationally state-selected HD beam at $r=0$ is considered.\label{fig11}}	
\end{figure}

\begin{figure}
	\includegraphics[width=1.0\columnwidth]{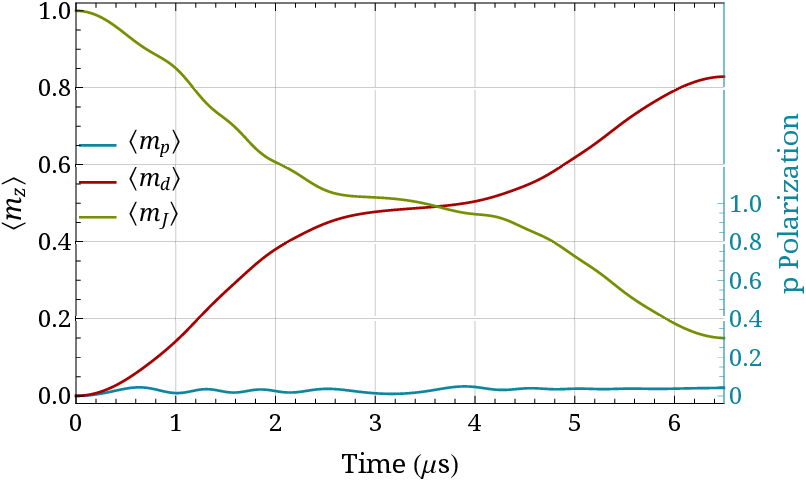}
	\caption{Average spin projections $\left\langle m_{p}\right\rangle$, $\left\langle m_{d}\right\rangle$, and $\left\langle m_{J}\right\rangle$ along the $z$-axis for $J = 1$ and $B_{0} = 48.6$~mT as a function of time. The right vertical axis represents the proton spin polarization, while the deuteron polarization is directly given by $\left\langle m_{d}\right\rangle$. An initially coherently rotationally state-selected HD beam at $r=0$ is considered.\label{fig12}}
\end{figure}

Figure~\ref{fig11} illustrates the variation of the average spin projections for the rotational level $J=1$ at the exit of the applied magnetic field $B_{z}$ with a frequency $1/t_{f} = 153935$~Hz, plotted as a function of the magnetic field amplitude $B_{0}$. The initial preparation assumes coherent excitation to states with $m_{J} = 1$, specifically the 6 states $\left \vert m_{J} = 1 ,\, m_{p} = \pm \frac{1}{2} , \, m_d = 0, \pm 1 \right\rangle$. The deuteron polarization reaches approximately $83\%$ at $B_{0}=48.6$~mT. Since the total number of states involved is quite large ($18$), making the state population dynamics complex, the spin projection evolution during the interaction time for this magnetic field amplitude is presented in Fig.~\ref{fig12} instead of the corresponding state populations.

\begin{figure}
	\includegraphics[width=1.0\columnwidth]{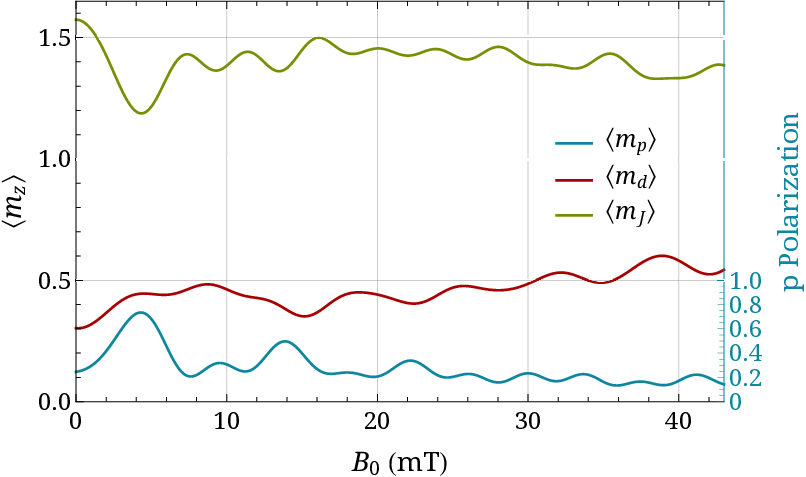}
	\caption{Average spin projections $\left\langle m_{p}\right\rangle$, $\left\langle m_{d}\right\rangle$, and $\left\langle m_{J}\right\rangle$ along the $z$-axis for $J = 2$ at $t_{f}\sim 11.4$~\textmu{s} as a function of $B_0$. The right vertical axis represents the proton spin polarization, while the deuteron polarization is directly given by $\left\langle m_{d}\right\rangle$. An initially coherently rotationally state-selected HD beam at $r=0$ is considered.\label{fig13}}
\end{figure}

\begin{figure}
	\includegraphics[width=1.0\columnwidth]{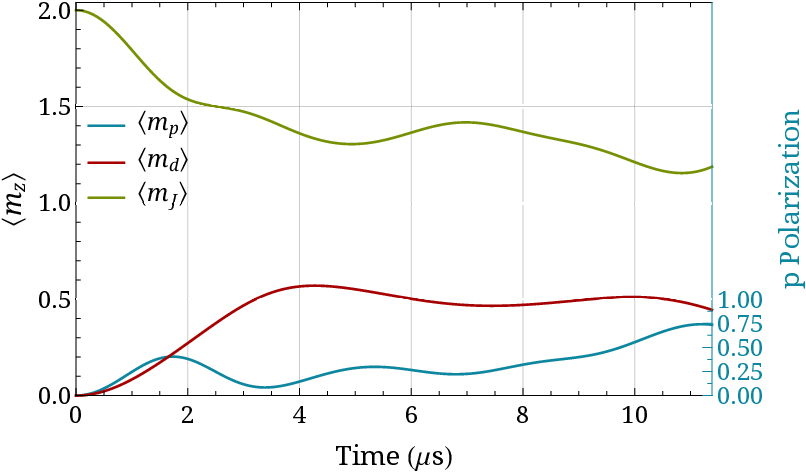}
	\caption{Average spin projections $\left\langle m_{p}\right\rangle$, $\left\langle m_{d}\right\rangle$, and $\left\langle m_{J}\right\rangle$ along the $z$-axis for $J = 2$ and $B_{0} = 4.3$~mT as a function of time. The right vertical axis represents the proton spin polarization, while the deuteron polarization is directly given by $\left\langle m_{d}\right\rangle$. An initially coherently rotationally state-selected HD beam at $r=0$ is considered.\label{fig14}}
\end{figure}

The same methodology is applied for $J=2$, focusing on enhancing proton polarization. Unlike $J=1$, where proton polarization can already reach high values (up to $70\%$) without external fields, $J=2$ requires such to reach similar values. Figure~\ref{fig13} displays the average spin projections for $B_{z}$ with $1/t_{f} = 87852$~Hz, considering an initial state prepared coherently and equally distributed among the $6$ states $\left \vert m_{J} = 2 ,\, m_{p} = \pm \frac{1}{2} , \, m_d = 0, \pm 1 \right\rangle$. The proton polarization peaks at $73.5\%$ for $B_{0}=4.3$~mT. The time evolution of the average projections for this magnetic field amplitude is shown in Fig.~\ref{fig14}. For this rotational level, the number of states increases to $30$.

An appendix section for HD is not provided, as there is no need to introduce a specific coupling scheme to facilitate the description of the initial states, and the large number of states makes writing out the matrix representation of operators impractical. The spin matrices required for the calculations of the average spin projections are obtained similarly to those for H and D atoms (see Apps.~\ref{app1a} and~\ref{app1b}). For example, $S_{z, d, u} = \mathds{1}_{J} \otimes \mathds{1}_p \otimes \hbar \sigma_{z, 1}$ which is represented by a diagonal matrix with elements $0$ and $\pm1$, each appearing with a multiplicity of $6$ for $J=1$. Here, $\mathds{1}_{J}$ and $\mathds{1}_p$ are the $3 \times 3$ and $2 \times 2$ identity matrices, respectively, while $\sigma_{z, 1}$ is the Pauli matrix for spin $1$, as defined in App.~\ref{app1b}.

For a $1$-keV HD beam ($v\sim 2.5\times 10^{5}$~m/s), the times of flight corresponding to the analyzed field frequencies translate to wavelengths of $1.6$~m and $2.9$~m. These are significantly longer than the (sub)millimeter wavelengths typical of atomic systems, reflecting the difference in energy (or frequency) scales of the hyperfine splittings. The effect of the radial field component on off-axis particles is analyzed for $r=2$~cm, rather than at radial distances comparable to $\lambda$. As previously explained, the influence of $B_{r}$ is minimal because $r\ll \lambda$. The slight loss of polarization at this radial distance for both wavelengths is demonstrated in App.~\ref{app2b} (see Figs.~\ref{fig15} and~\ref{fig16}) while varying the magnetic field amplitude $B_{0}$. If, instead, velocities of $1$~km/s are adopted, e.g., from supersonic beam expansion, wavelengths of a few centimeters can be used.

\subsection{Limitations\label{limitations}}  

The theoretical framework developed in this work describes a purely magnetic spin model that simulates ensembles of non-interacting particles exposed to a time-dependent magnetic field of the form given in Eq.~(\ref{eq:bfieldt}), while disregarding interactions with additional fields, such as electric fields, that may also be present. It is devised as an equivalent picture for the spin dynamics of a particle beam moving with a constant non-relativistic velocity $v$ through a static, spatially varying magnetic field. The transformation between the two reference frames assumes that the time coordinate is the same for both. Such a transformation is known as a Galilean transformation and is an approximation of the Lorentz transformation in the limit of relative speeds $v$ much less than the speed of light $c$ in vacuum. However, the framework can be expanded to relativistic beam velocities by incorporating the necessary modifications dictated by a Lorentz transformation. Below, we give the restrictions on the beam kinetic energy range for which this approximation is valid.

Conventionally, relativistic effects are ignored when $\beta {=} v/c {<} 1\%$, i.e., $v{<}2.998{\times} 10^{6}$~m/s. This sets an upper limit on the kinetic energy for beams of H, D, or HD, at $46.9$~keV, $93.8$~keV, and $140.7$~keV, respectively. These limits cover a wide range of experimental setups in which ions are accelerated electrostatically before forming neutral systems, such as atoms or molecules.

As mentioned above, the model introduced for evaluating spin dynamics assumes purely magnetic interactions. However, transitioning from the laboratory frame to the beam's rest frame introduces an additional electric field due to the relative motion of the particle beam. This field is described in the non-relativistic limit by $\mathbf{E} =\mathbf{v}\times\mathbf{B}$~\cite{galili1997, koenig2021}. The phenomenon of the interaction of the system with such a field is known as the motional Stark effect~\cite{levinton1999}.  It arises from the radial magnetic field component $B_{r}$ (i.e., when $r\neq 0$). Since the beam velocity is parallel to the $z$-axis, the electric field lies in the transverse plane,
\begin{equation}
	\mathbf{E} = v B_{r} \, \hat{\boldsymbol{\phi}} = v B_{r} (-\sin\phi \, \hat{\mathbf{x}} + \cos\phi \, \hat{\mathbf{y}}) . 
\end{equation}
The amplitude of this field scales with the beam velocity and the radial magnetic field component.

The interaction Hamiltonian is given by
\begin{equation}\label{eq:hE}
	H_{E} = - \mathbf{d} \cdot \mathbf{E} ,
\end{equation}
where $\mathbf{d}$ is the electric dipole moment of the system. This interaction does not couple states within the hyperfine regime considered here, but states with opposite parity. For example, for hydrogen and deuterium atoms the closest opposite-parity states to the ground state $1S_{1/2}$ or the metastable $2S_{1/2}$ are the $2P_{1/2}$ and $2P_{3/2}$ levels. A rough estimate of the energy difference that this interaction can cover is given by the quantity $e v B_{r} a_0$, where $e$ is the elementary charge and $a_0$ is the Bohr radius. For $v\sim 10^{5}$~m/s and $B_{r}\sim 1{-}100$~mT, this energy ranges from $5\times 10^{-9}$ to $5\times 10^{-7}$~eV. For comparison, the energy differences $1S_{1/2} - 2 P$ and $2S_{1/2}-2P_{1/2}$ for H are on the order of $10$ eV and a few \textmu{eV}, respectively. A detailed calculation of this effect would require an expanded basis, incorporating higher orbitals beyond the hyperfine regime of single orbitals examined here. An example of such calculations for metastable H can be found in~\cite{faatz2023,faatz2025}.

\begin{figure}
	\includegraphics[width=1.0\columnwidth]{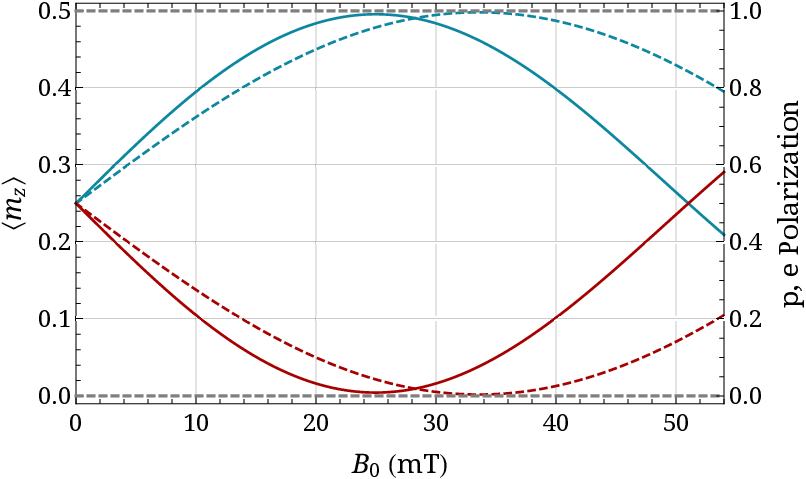}
	\caption{Comparison of average spin projections $\left\langle m_{z, p}\right\rangle$ (blue) and $\left\langle m_{z, e}\right\rangle$ (red) for a ground-state H atom under two magnetic field shapes: sine (solid line) and sine cubed (dashed line). The right vertical axis represents the corresponding polarization.\label{fig17}}
\end{figure}

\begin{figure}  
	\includegraphics[width=1.0\columnwidth]{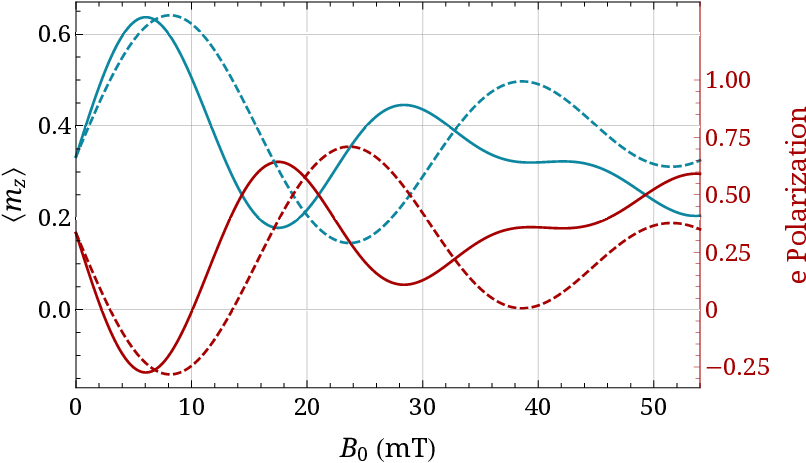}
	\caption{Comparison of average spin projections $\left\langle m_{z, d}\right\rangle$ (blue) and $\left\langle m_{z, e}\right\rangle$ (red) for a ground-state D atom under two magnetic field shapes: sine (solid line) and sine cubed (dashed line). The right vertical axis represents the electron spin polarization, while the deuteron polarization is directly given by $\left\langle m_{z, d}\right\rangle$.\label{fig18}}
\end{figure}

Moreover, while this work uses analytically defined magnetic field configurations for clarity, exact adherence to such configurations is not strictly necessary for the effective application of spin-enhancement techniques. By tuning the magnetic field amplitude and/or time of flight, high nuclear polarization can still be achieved even with non-ideal field shapes. To demonstrate this, Figs.~\ref{fig17} and~\ref{fig18} compare the results for a longitudinal field $B_{z}$ proportional to a sine function (see Figs.~\ref{fig1} and~\ref{fig5}) versus a sine-cubed function, for H and D, respectively. The dashed lines correspond to the sine-cubed magnetic field configuration. The nuclear polarization peaks at different field amplitudes $B_{0}$ and reaches slightly higher values, $99.6\%$ for the protons and $64.1\%$ for the deuterons.

The applicability of this study also depends on achieving and maintaining the discussed initial state preparations. For atomic systems, where initial partial polarization is assumed, this can be sustained by minimizing depolarization effects. The same applies to molecular systems, but an additional factor is the coherence involved in the presumed excitation. A typical example is stimulated Raman adiabatic passage (STIRAP)~\cite{vitanov2017}, which is susceptible to decoherence effects. The dominant source of decoherence is phase relaxation, which can be mitigated by reducing the pulse width and/or increasing the pulse delay while maintaining adiabaticity~\cite{ivanov2004}. Decoherence also arises from interactions between the system and its external environment. However, for the timescales considered here, preserving coherence is not expected to be a significant challenge.

Finally, the flexibility of the theoretical framework extends to the ability to incorporate various magnetic field configurations as input. While this paper uses analytical expressions to facilitate discussions, the numerical evaluation of spin dynamics readily accommodates arbitrary field configurations provided as data points or interpolated functions, as in~\cite{engels2021}.

\section{Conclusion}

We have developed a versatile theoretical and computational framework for analyzing the spin dynamics of non-relativistic particle beams with interacting angular momenta in static, spatially varying magnetic fields. The considered field configurations are inspired by the experimental realization of Sona's proposal for polarized ion sources in accelerator physics~\cite{sona1967}. Notably, this approach also extends to stationary systems under time-dependent magnetic fields, provided the initial state is a statistical mixture.

A particularly promising application of this framework is its potential to enhance nuclear polarization for the production of polarized fuel for fusion reactors. Our studies on hydrogen, deuterium, and hydrogen deuteride demonstrate that a single, adaptable spin manipulation scheme can significantly boost nuclear polarization. Future work will extend this analysis to tritium, likely within the tritium deuteride molecule, whose hyperfine structure has been extensively calculated~\cite{jozwiak2021}. In addition, the proposed technique can be adapted to enhance tensor polarization in deuterium. Further spin manipulations are also possible, not necessarily limited to nuclear spin, depending on the desired spin configuration. 

Moreover, the developed framework can be used to evaluate the spin dynamics in particle beams passing through arbitrarily varying magnetic fields. This includes fields that do not necessarily cross zero or change direction, as well as transitions between regions of homogeneous fields in setups for storing spin-polarized particles~\cite{kannis2021, kannis2022}. Magnetic fields with structures ranging from submillimeter scales~\cite{nagata2017} to meter scales~\cite{chadwick2022} are commonly employed in experiments, both of which are encompassed by the atomic and molecular systems discussed here. In the case of periodic fields, the Fourier transform of time-domain observables introduced in this work enables the prediction of observed signals in magnetic resonance experiments~\cite{nagata2017}. This feature broadens the framework's applicability, offering utility in diverse experimental scenarios and advancing our understanding of spin dynamics in complex systems.

\begin{acknowledgments}
This work is funded by the Deutsche Forschungsgemeinschaft (DFG, German Research Foundation)-533904660.
\end{acknowledgments}

\appendix

\section{Coupled and uncoupled representations\label{app1}}

In the following, the coupled and uncoupled representations are described for the H and D atoms, for which the total electron angular momentum equals its spin. The transformation between these bases is presented and the matrix representations of the spin components (spin matrices) are shown. To maintain compatibility with standard conventions, the quantization axis is considered to be the $z$-axis.

\subsection{System of two spin-1/2 particles\label{app1a}}

As described in Sec.~\ref{theory}, a system of two angular momenta with $L_{1,2} = \frac{1}{2}$ can be described by the basis vectors $\left\vert l_{1}, l_{2}\right\rangle$, where $l_{1,2}$ are the projections along the quantization axis, taking values $l_{1,2}=\pm \frac{1}{2}$. The four elements $\left\vert l_{1}, l_{2}\right\rangle$ compose the uncoupled representation (note that the notation will change from the generic $L_{1,2}$ and $l_{1,2}$ to $S,\, I$ and $m_{S,I}$ to be compatible with the spin operators in Sec.~\ref{resndisc}). The elements of the basis are arranged in the following sequence: 
\begin{equation}
	\begin{split}
		\text{uncoupled basis $\{\left\vert m_{S}, m_{I}\right\rangle\}$: } &\Bigg\{\left\vert  \frac{1}{2} , \frac{1}{2} \right\rangle , \,  \left\vert \frac{1}{2} , -\frac{1}{2} \right\rangle ,  \\
		& \left\vert -\frac{1}{2} , -\frac{1}{2} \right\rangle , \, \left\vert -\frac{1}{2} , \frac{1}{2} \right\rangle\Bigg\}
	\end{split}
\end{equation}

Adding $\mathbf{S}$ and $\mathbf{I}$, the resultant angular momentum $	\mathbf{F} = \mathbf{S} \otimes \mathds{1} + \mathds{1} \otimes \mathbf{I}$ has three possible projections for $F= 1$ (triplet) and one for $F = 0$ (singlet). Therefore, the coupled basis is given by (again, note that in contrast to Sec.~\ref{theory}, where the generic letters $K,\, k$ were utilized, here we use the notation $F,\, m_{F}$ according to textbook nomenclature):
\begin{equation}
	\text{coupled basis $\{\left\vert F, m_{F}\right\rangle\}$: } \big\{\left\vert 1, 1 \right\rangle , \, \left\vert 1, 0 \right\rangle , \, \left\vert 1, -1 \right\rangle , \, \left\vert 0, 0 \right\rangle \big\} 
\end{equation}

The relationships between the kets of the two representations are determined using the Clebsch-Gordan coefficients:
\begin{subequations}
	\begin{flalign}\label{eq:c1}
		&\left\vert F = 1, m_{F} = 1 \right\rangle = \left\vert m_{S} = \frac{1}{2} , m_{I} = \frac{1}{2} \right\rangle &\\ \label{eq:c2}
		&\left\vert 1, 0 \right\rangle = \frac{1}{\sqrt{2}} \Bigg( \left\vert \frac{1}{2} , -\frac{1}{2} \right\rangle + \left\vert -\frac{1}{2} , \frac{1}{2} \right\rangle \Bigg) &\\ \label{eq:c3}
		&\left\vert 1, -1 \right\rangle = \left\vert -\frac{1}{2} , -\frac{1}{2} \right\rangle &\\ \label{eq:c4}
		&\left\vert 0, 0 \right\rangle = \frac{1}{\sqrt{2}} \Bigg( \left\vert \frac{1}{2} , -\frac{1}{2} \right\rangle - \left\vert -\frac{1}{2} , \frac{1}{2} \right\rangle \Bigg)
	\end{flalign}
\end{subequations}
If we denote the vector states on the left-hand side by $\left\vert \chi \right\rangle_{c}$ and those on the right-hand side by $\left\vert \chi \right\rangle_{u}$, Eqs.~(\ref{eq:c1})--(\ref{eq:c4}) can be compactly expressed as
\begin{equation}\label{eq:cu}
	\left\vert \chi \right\rangle_{c} = \mathcal{Q} \left\vert \chi \right\rangle_{u} \quad \text{with} \quad
	\mathcal{Q} = 
	\begin{pmatrix}
		1 & 0 & 0 & 0 \\
		0 & \frac{1}{\sqrt{2}} & 0 & \frac{1}{\sqrt{2}} \\
		0 & 0 & 1 & 0 \\
		0 & \frac{1}{\sqrt{2}} & 0 & -\frac{1}{\sqrt{2}}
	\end{pmatrix},
\end{equation}
where $\mathcal{Q}$ is the transformation matrix that transforms a state vector expressed in the uncoupled basis into its representation in the coupled basis. It is an orthonormal matrix, so the inverse transformation 
\begin{equation}
\left\vert \chi \right\rangle_{u} = \mathcal{Q}^{-1} \left\vert \chi \right\rangle_{c}
\end{equation}
is given by the transpose of the matrix $\mathcal{Q}$, i.e., $\mathcal{Q}^{-1} = \mathcal{Q}^{T}$.

Operators also transform between the two representations as follows:
\begin{align}\label{eq:cu2}
	&	\mathcal{T}_c = \mathcal{Q} \, \mathcal{T}_u \, \mathcal{Q}^{-1} , \\
	\label{eq:cut2}
	&	\mathcal{T}_u   = \mathcal{Q}^{-1} \, \mathcal{T}_c \, \mathcal{Q} ,
\end{align}
where $\mathcal{T}_{c,u}$ are the operators in the coupled and uncoupled basis, respectively. 

Using the Pauli matrices $\sigma_{q,1/2}$ for $q=x,y,z$,
\begin{equation}\label{eq:paulisigma1}
\sigma_{x,1/2} =
\begin{pmatrix}
	0 & 1  \\
	1 & 0  
\end{pmatrix}
, \,
\sigma_{y,1/2} =
\begin{pmatrix}
	0 & -i  \\
	i & 0  
\end{pmatrix}
, \,
\sigma_{z,1/2} =
\begin{pmatrix}
	1 & 0  \\
	0 & -1  
\end{pmatrix}
,
\end{equation}
we derive the spin matrices for the proton (subscript $p$) and electron (subscript $e$) in the uncoupled representation according to
\begin{equation}
	S_{q, p, u} = \mathds{1}_e \otimes \frac{\hbar}{2} \sigma_{q, 1/2} \quad \text{and} \quad S_{q, e, u} = \frac{\hbar}{2} \sigma_{q, 1/2} \otimes \mathds{1}_p  ,
\end{equation}
where
\begin{equation}
	\mathds{1}_e=\mathds{1}_p =
	\begin{pmatrix}
		1 & 0  \\
		0 & 1  
	\end{pmatrix}
	.
\end{equation}

The spin matrices in the uncoupled representations are listed below:
\begin{equation}\label{eq:paulim1a}
	\begin{aligned}
		& 	S_{x, p, u} = \frac{\hbar}{2}
		\begin{pmatrix}
			0 & 1 & 0 & 0 \\
			1 & 0 & 0 & 0 \\
			0 & 0 & 0 & 1 \\
			0 & 0 & 1 & 0 
		\end{pmatrix}
		, \,
		S_{x, e, u} = \frac{\hbar}{2}
		\begin{pmatrix}
			0 & 0 & 0 & 1 \\
			0 & 0 & 1 & 0 \\
			0 & 1 & 0 & 0 \\
			1 & 0 & 0 & 0 
		\end{pmatrix}  \\ %\label{eq:paulim1b}
		&	S_{y, p, u} = \frac{\hbar}{2}
		\begin{pmatrix}
			0 & -i & 0 & 0 \\
			i & 0 & 0 & 0 \\
			0 & 0 & 0 & i \\
			0 & 0 & -i & 0 
		\end{pmatrix}
		, \,
		S_{y, e, u} = \frac{\hbar}{2}
		\begin{pmatrix}
			0 & 0 & 0 & -i \\
			0 & 0 & -i & 0 \\
			0 & i & 0 & 0 \\
			i & 0 & 0 & 0 
		\end{pmatrix} \\ %\label{eq:paulim1c}
		&	S_{z, p, u} = \frac{\hbar}{2}
		\begin{pmatrix}
			1 & 0 & 0 & 0 \\
			0 & -1 & 0 & 0 \\
			0 & 0 & -1 & 0 \\
			0 & 0 & 0 & 1 
		\end{pmatrix}			
		, \,
		S_{z, e, u} = \frac{\hbar}{2}
		\begin{pmatrix}
			1 & 0 & 0 & 0 \\
			0 & 1 & 0 & 0 \\
			0 & 0 & -1 & 0 \\
			0 & 0 & 0 & -1 
		\end{pmatrix}  
		.
	\end{aligned}
\end{equation}
As an example, we calculate the proton spin matrix in the coupled representation, combining Eq.~(\ref{eq:paulim1a}) and Eq.~(\ref{eq:cu2}),
\begin{equation}\label{eq:ex1}
	S_{z, p, c} = \mathcal{Q} \, S_{z, p, u} \, \mathcal{Q}^{-1} = \frac{\hbar}{2}
	\begin{pmatrix}
		1 & 0 & 0 & 0 \\
		0 & 0 & 0 & -1 \\
		0 & 0 & -1 & 0 \\
		0 & -1 & 0 & 0 
	\end{pmatrix}.
\end{equation}
Likewise, the other spin matrices can be obtained in the coupled representation.

\subsection{System of a spin-1/2 and a spin-1 particle}\label{app1b}

The electron and deuteron spin quantum numbers, considered here, are $1/2$ and $1$, respectively. The same procedure as in App.~\ref{app1a} is followed. The uncoupled basis comprises of the following vector states,
\begin{equation}
	\begin{split}
		\{\left\vert m_{S}, m_{I}\right\rangle\} : 
		&\Bigg\{\left\vert  \frac{1}{2} , 1 \right\rangle , \,  \left\vert \frac{1}{2} , 0 \right\rangle ,  \,  \left\vert \frac{1}{2} , -1 \right\rangle ,  \\
		& \left\vert -\frac{1}{2} , -1 \right\rangle , \,  \left\vert -\frac{1}{2} , 0 \right\rangle, \, \left\vert -\frac{1}{2} , 1 \right\rangle\Bigg\},
	\end{split}
\end{equation}
while the coupled basis consists of
\begin{equation}
	\begin{split}
		\{\left\vert F, m_{F}\right\rangle\}: 
		&\Bigg\{\left\vert \frac{3}{2}, \frac{3}{2} \right\rangle , \, \left\vert \frac{3}{2}, \frac{1}{2} \right\rangle , \, \left\vert \frac{3}{2}, -\frac{1}{2} \right\rangle , \, \left\vert \frac{3}{2}, -\frac{3}{2} \right\rangle, \\
		& \left\vert \frac{1}{2}, -\frac{1}{2} \right\rangle , \, \left\vert \frac{1}{2}, \frac{1}{2} \right\rangle \Bigg\} .
	\end{split}
\end{equation}

The Clebsch-Gordan decomposition yields
\begin{subequations}
	\begin{flalign}\label{eq:c1b}
		&\left\vert F = \frac{3}{2}, m_{F} = \frac{3}{2} \right\rangle = \left\vert m_{S} = \frac{1}{2} , m_{I} = 1 \right\rangle &\\ \label{eq:c2b}
		&\left\vert \frac{3}{2}, \frac{1}{2} \right\rangle = \sqrt{\frac{2}{3}}  \left\vert \frac{1}{2} , 0 \right\rangle + \frac{1}{\sqrt{3}} \left\vert -\frac{1}{2} , 1 \right\rangle  &\\ \label{eq:c3b}
		&\left\vert \frac{3}{2}, -\frac{1}{2} \right\rangle = \frac{1}{\sqrt{3}} \left\vert \frac{1}{2} , -1 \right\rangle + \sqrt{\frac{2}{3}} \left\vert -\frac{1}{2} , 0 \right\rangle &\\ \label{eq:c4b}
		&\left\vert \frac{3}{2}, -\frac{3}{2} \right\rangle = \left\vert -\frac{1}{2} , -1 \right\rangle &\\ \label{eq:c5b}
		&\left\vert \frac{1}{2}, -\frac{1}{2} \right\rangle = \sqrt{\frac{2}{3}}  \left\vert \frac{1}{2} , -1 \right\rangle - \frac{1}{\sqrt{3}} \left\vert -\frac{1}{2} , 0 \right\rangle  &\\ \label{eq:c6b}
		&\left\vert \frac{1}{2}, \frac{1}{2} \right\rangle = \frac{1}{\sqrt{3}}  \left\vert \frac{1}{2} , 0 \right\rangle -\sqrt{\frac{2}{3}}  \left\vert -\frac{1}{2} , 1 \right\rangle
	\end{flalign}
\end{subequations}
Similar to App.~\ref{app1a} the orthonormal transformation matrix $\mathcal{Q}$ is expressed as
\begin{equation}\label{eq:cub}
	\mathcal{Q} = 
	\begin{pmatrix}
		1 & 0 & 0 & 0 & 0 & 0 \\
		0 & \sqrt{\frac{2}{3}} & 0 & 0 & 0 & \frac{1}{\sqrt{3}} \\
		0 & 0 & \frac{1}{\sqrt{3}} & 0 & \sqrt{\frac{2}{3}} & 0 \\
		0 & 0 & 0 & 1 & 0 & 0 \\
		0 & 0 & \sqrt{\frac{2}{3}} & 0 & -\frac{1}{\sqrt{3}} &0 \\
		0 & \frac{1}{\sqrt{3}} & 0 & 0 & 0 & -\sqrt{\frac{2}{3}}
	\end{pmatrix}
\end{equation}
and the basis transformations follow Eq.~(\ref{eq:cu}) for vector states and Eqs.~(\ref{eq:cu2}) and~(\ref{eq:cut2}) for operators.

The Pauli matrices for spin $1$, denoted by $\sigma_{q, 1}$ with $q=x,y,z$ are
\begin{align}\label{eq:paulisigma2}
 \begin{aligned}
	&\begin{array}{c}
	\sigma_{x, 1} = \frac{1}{\sqrt{2}}
	\begin{pmatrix}
		0 & 1 & 0  \\
		1 & 0 & 1 \\
		0 & 1 & 0
	\end{pmatrix}
	, \quad
	\sigma_{y, 1} = \frac{1}{\sqrt{2}}
	\begin{pmatrix}
		0 & -i & 0  \\
		i & 0  & -i \\
		0 & i & 0
	\end{pmatrix}
	, \\
	\sigma_{z, 1} =
	\begin{pmatrix}
		1 & 0 & 0  \\
		0 & 0 & 0  \\
		0 & 0 & -1  
	\end{pmatrix}
	.
\end{array}
\end{aligned}
\end{align}

The spin matrices for the deuteron and electron are obtained as follows:
\begin{equation}
	S_{q, d, u} = \mathds{1}_e \otimes \hbar \sigma_{q, 1} \quad \text{and} \quad S_{q, e, u} = \frac{\hbar}{2} \sigma_{q, 1/2} \otimes \mathds{1}_d  ,
\end{equation}
where
\begin{equation}
	\mathds{1}_d =
	\begin{pmatrix}
		1 & 0 & 0  \\
		0 & 1  & 0 \\
		0 & 0 & 1
	\end{pmatrix},
\end{equation}
and $\mathds{1}_e$ and $\sigma_{q, 1/2}$ are $2 \times 2$ matrices given in App.~\ref{app1a}. 

\begin{widetext}
The matrices in the uncoupled representation are summarized:
	\begin{equation}\label{eq:paulim1b}
		\begin{aligned}
			& 	S_{x, d, u} = \frac{\hbar}{\sqrt{2}}
			\begin{pmatrix}
				0 & 1 & 0 & 0 & 0 & 0 \\
				1 & 0 & 1 & 0 & 0 & 0 \\
				0 & 1 & 0 & 0 & 0 & 0 \\
				0 & 0 & 0 & 0 & 1 & 0 \\
				0 & 0 & 0 & 1 & 0 & 1 \\
				0 & 0 & 0 & 0 & 1 & 0
			\end{pmatrix}
			, \quad
			S_{y, d, u} = \frac{\hbar}{\sqrt{2}}
			\begin{pmatrix}
				0 & -i & 0 & 0 & 0 & 0 \\
				i & 0 & -i & 0 & 0 & 0 \\
				0 & i & 0 & 0 & 0 & 0 \\
				0 & 0 & 0 & 0 & i & 0 \\
				0 & 0 & 0 & -i & 0 & i \\
				0 & 0 & 0 & 0 & -i & 0
			\end{pmatrix}
			, \quad
			S_{z, d, u} = \hbar
			\begin{pmatrix}
				1 & 0 & 0 & 0 & 0 & 0 \\
				0 & 0 & 0 & 0 & 0 & 0 \\
				0 & 0 & -1 & 0 & 0 & 0 \\
				0 & 0 & 0 & -1 & 0 & 0 \\
				0 & 0 & 0 & 0 & 0 & 0 \\
				0 & 0 & 0 & 0 & 0 & 1
			\end{pmatrix}
			\\ 
			&	S_{x, e, u} = \frac{\hbar}{2}
			\begin{pmatrix}
				0 & 0 & 0 & 0 & 0 & 1 \\
				0 & 0 & 0 & 0 & 1 & 0 \\
				0 & 0 & 0 & 1 & 0 & 0 \\
				0 & 0 & 1 & 0 & 0 & 0 \\
				0 & 1 & 0 & 0 & 0 & 0 \\
				1 & 0 & 0 & 0 & 0 & 0
			\end{pmatrix}
			, \quad
			S_{y, e, u} = \frac{\hbar}{2}
			\begin{pmatrix}
				0 & 0 & 0 & 0 & 0 & -i \\
				0 & 0 & 0 & 0 & -i & 0 \\
				0 & 0 & 0 & -i & 0 & 0 \\
				0 & 0 & i & 0 & 0 & 0 \\
				0 & i & 0 & 0 & 0 & 0 \\
				i & 0 & 0 & 0 & 0 & 0
			\end{pmatrix} 
			, \quad
			S_{z, e, u} = \frac{\hbar}{2}
			\begin{pmatrix}
				1 & 0 & 0 & 0 & 0 & 0 \\
				0 & 1 & 0 & 0 & 0 & 0 \\
				0 & 0 & 1 & 0 & 0 & 0 \\
				0 & 0 & 0 & -1 & 0 & 0 \\
				0 & 0 & 0 & 0 & -1 & 0 \\
				0 & 0 & 0 & 0 & 0 & -1 
			\end{pmatrix}  
			.
		\end{aligned}
	\end{equation}

The calculation of the $z$-component of the deuteron spin matrix is given below as an example of the basis transformation application:
	\begin{equation}\label{eq:ex2}
		S_{z, d, c} = \mathcal{Q} \, S_{z, d, u} \, \mathcal{Q}^{-1} = \hbar
		\begin{pmatrix}
			1 & 0 & 0 & 0 & 0 & 0 \\
			0 & \frac{1}{3} & 0 & 0 & 0 & -\frac{\sqrt{2}}{3} \\
			0 & 0 & -\frac{1}{3} & 0 & -\frac{\sqrt{2}}{3} & 0 \\
			0 & 0 & 0 & -1 & 0 & 0 \\
			0 & 0 & -\frac{\sqrt{2}}{3} & 0 & -\frac{2}{3} & 0 \\
			0 & -\frac{\sqrt{2}}{3} & 0 & 0 & 0 & \frac{2}{3}
		\end{pmatrix}.
	\end{equation}
Likewise, the other spin matrices can be obtained in the coupled representation. 
\end{widetext}

\section{Supplementary figures\label{app2}}

Additional visualizations for D and HD are presented here to complement the main results discussed in this work.

\subsection{Deuterium plots\label{app2a} }

\begin{figure}[h]
	\includegraphics[width=1.0\columnwidth]{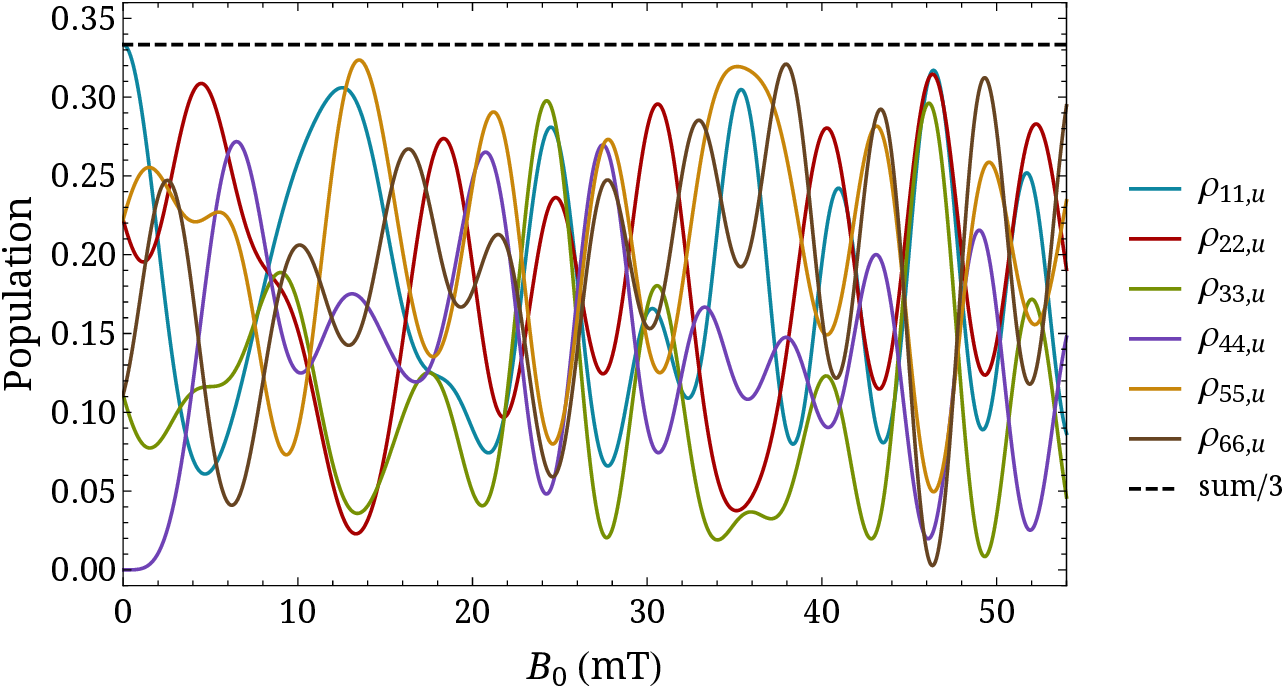}
	\caption{Populations of uncoupled states at time $t_{f}\sim 3$~ns as a function of $B_0$, at $r\sim 0.9$~mm. The states are ordered according to App.~\ref{app1b}.\label{fig9}}
\end{figure}
\begin{figure}[h]
	\includegraphics[width=1.0\columnwidth]{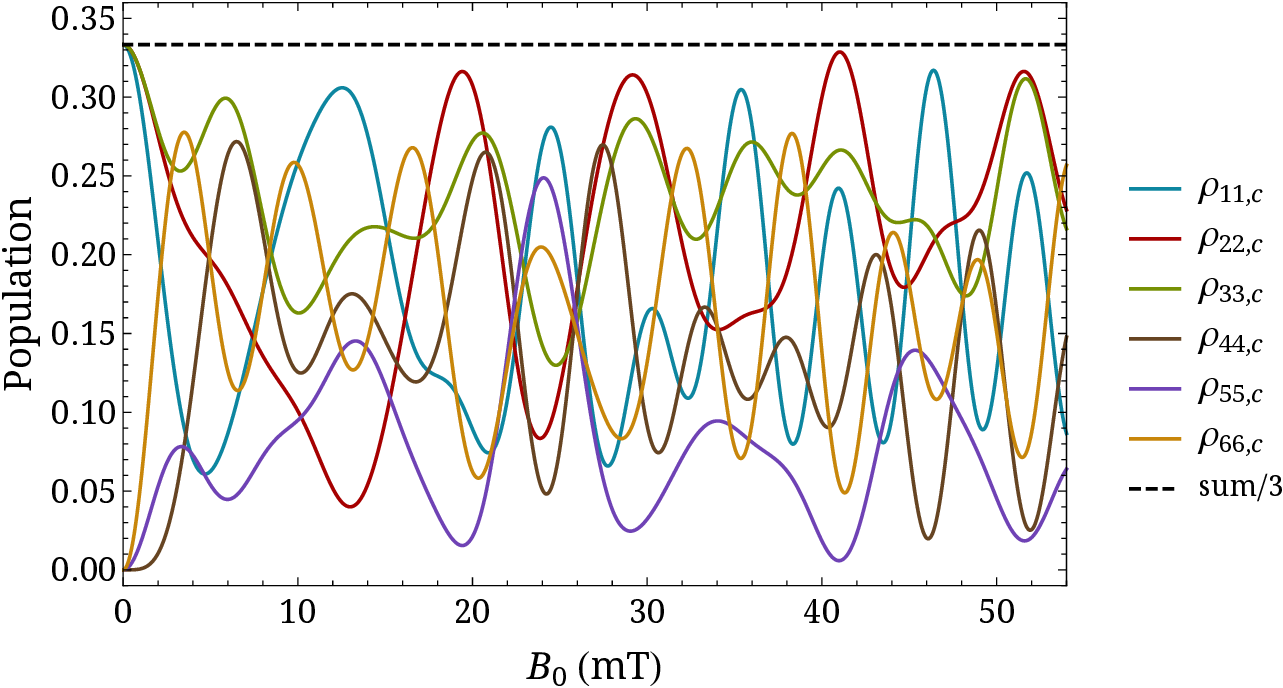}
	\caption{Populations of coupled states at time $t_{f}\sim 3$~ns as a function of $B_0$, at $r\sim 0.9$~mm. The states are ordered according to App.~\ref{app1b}.\label{fig10}}
\end{figure}

\subsection{Hydrogen deuteride plots\label{app2b}}

\begin{figure}[h]
	\includegraphics[width=0.87\columnwidth]{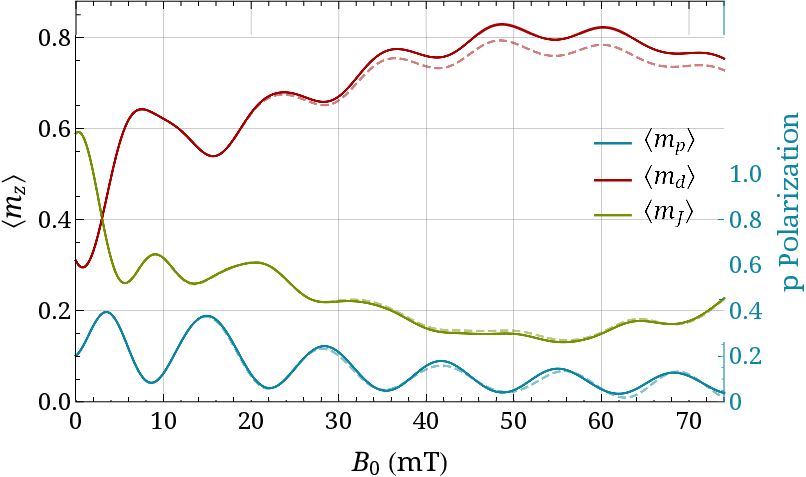}
	\caption{Average spin projections $\left\langle m_{p}\right\rangle$, $\left\langle m_{d}\right\rangle$, and $\left\langle m_{J}\right\rangle$ along the $z$-axis for $J = 1$ at $t_{f}\sim 6.5$~\textmu{s} as a function of $B_0$. The right vertical axis represents the proton spin polarization, while the deuteron polarization is directly given by $\left\langle m_{d}\right\rangle$. An initially coherently rotationally state-selected HD beam at $r=0$ (solid lines) and  $r=2$~cm (dashed lines) is considered.\label{fig15}}
\end{figure}
\begin{figure}[h]
	\includegraphics[width=0.87\columnwidth]{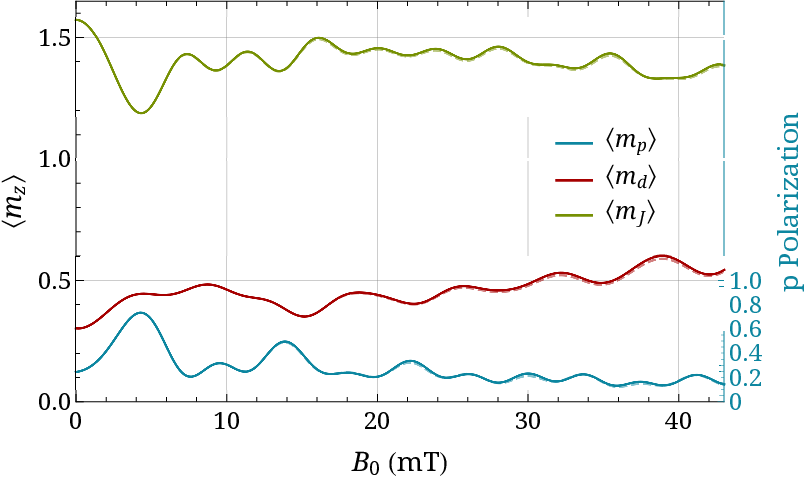}
	\caption{Average spin projections $\left\langle m_{p}\right\rangle$, $\left\langle m_{d}\right\rangle$, and $\left\langle m_{J}\right\rangle$ along the $z$-axis for $J = 2$ at $t_{f}\sim 11.4$~\textmu{s} as a function of $B_0$. The right vertical axis represents the proton spin polarization, while the deuteron polarization is directly given by $\left\langle m_{d}\right\rangle$. An initially coherently rotationally state-selected HD beam at $r=0$ (solid lines) and  $r=2$~cm (dashed lines) is considered.\label{fig16}}
\end{figure}

%\newpage
%\clearpage

%\bibliography{pra2025_kannis_biblio}

%

\end{document}